# High-Speed Imagery Analysis of Droplet Impact on Van der Waals and Non-Van der Waals Soft-Textured Oil-Infused Surfaces.

Shubham S. Ganar,[a] Deepak J.,[a] and Arindam Das*[a]

[1]*School of Mechanical Sciences, Indian Institute of Technology (IIT) Goa, GEC Campus, Farmagudi, Ponda, Goa 403401, India*

## Abstract

Droplet impact on Slippery Liquid-Infused Porous Surfaces has garnered significant attention due to its scientific importance and wide range of industrial applications. This study investigates the impact of surface functionalization, oil coating, and oil absorption on droplet impact behavior on textured polydimethylsiloxane(PDMS) substrates. The textured surfaces were fabricated with square micro-posts having spacings of 5μm and 20μm. The PDMS was functionalized with octadecyltrichlorosilane(OTS) to reduce surface energy and improve water repellency. Following functionalization, the surfaces were either coated with or allowed to absorb two different lubricants: 5cSt silicone oil (SO-5cSt) and hexadecane. High-speed imaging was then used to capture droplet impact across a wide range of Weber numbers. On SO-5cSt absorbed substrates, droplets showed complete rebound at all Weber numbers, regardless of post spacing. This robust rebound was attributed to the oil's ability to fill the gaps between the posts through capillary action, while also forming a stable lubricating layer above the texture. This thin oil film reduced friction between the droplet and the surface, enabling the droplet to rebound. In contrast, hexadecane-absorbed substrates displayed different dynamics. At low Weber numbers for 20μm, only partial rebound was observed, while at intermediate values, droplets rebounded completely. The repeated droplet-impact experiment further demonstrated that hexadecane-infused surfaces gradually lost oil from the textured gaps, leading to a decline in rebound performance over time. This effect was not observed with SO-5cSt, underscoring the importance of retention. Overall, this work demonstrates that surface functionalization, geometry, and lubricant properties collectively determine droplet mobility.

**Corresponding Author**

Arindam Das*, Associate Professor, School of Mechanical Sciences, Indian Institute of Technology (IIT) Goa, Email: arindam@iitgoa.ac.in,

# 1. Introduction

The interaction of liquid droplets with solid surfaces has fascinated scientists and engineers since Worthington's pioneering observations in 1876.[1] It began as a curiosity about the complex, crown-like shapes produced upon impact, and has evolved into a critical research area with wide-ranging technological implications.[1] Droplet impact dynamics, including spreading, retraction, rebound, and breakup, directly influence processes such as agricultural spraying, spray cooling, surface coating, inkjet printing, combustion, and even forensic investigations[2–5]. When a droplet strikes a rigid surface, it undergoes deformation, entrains a thin air pocket, and, at sufficiently high impact velocities, produces splashing that may either form a crown (corona splashing) or eject secondary droplets directly from the spreading lamella (prompt splashing).[6–8]The distinction between these splashing modes has significant implications in fields where the trajectory and size distribution of satellite droplets determine system efficiency, safety, such as fuel combustion[9] temperature dependences[10] and the transmission of pathogens.[11] Despite decades of study, the physics governing droplet impact remains incomplete, as modern high-speed imaging continues to reveal unanticipated features of this seemingly simple yet complex phenomenon.

Superhydrophobic surfaces, which trap air in their microstructures, have long been studied for their water-repellent properties.[6,12,13] When a water droplet is gently placed on a micropatterned surface, it typically settles into either the Cassie–Baxter or Wenzel wetting state. Patankar[14,15] developed a theoretical framework to describe the infiltration of liquids into the gaps between micropillars. Building on this, Nonomura et al.[16] used high-speed imaging to capture the moment a water droplet entered a pore on a silicone surface. An investigation by Bharat Bhushan et al.[17] focused on the effects of wetting behavior on droplet impact. In their study,[17] they observed that higher droplet impact velocities on silicon micropillars and PMMA surfaces led to a transition from a non-wetting (Cassie–Baxter) state to a wetting (Wenzel) state, whereas the MWCNT surface allowed the droplet to bounce off even at higher velocities. These findings are essential for designing superhydrophobic surfaces that retain water-repellent behavior under real-world, dynamic conditions. Researchers have fabricated a variety of soft, textured surfaces using materials such as PDMS to study droplet impact behaviour on superhydrophobic surfaces.[18] Ying-Song Yu et al.[19] conducted a droplet impact study on PDMS surfaces with post arrays of varying solid fractions. The results showed that surfaces with lower solid fractions allowed droplet rebound only at lower impact velocities. When the Weber number exceeded a critical threshold, a transition from the Cassie–Baxter to the Wenzel wetting state was observed. A predictive model was developed to capture this behavior, and the maximum spreading was found to follow a $We^{0.25}$ scaling law. Similar studies [20,21] investigated droplet impact on PDMS surfaces with microgrooves and different post spacings, shapes, and softness. This work systematically explored the effects of mechanical softness and microtexture on the dynamics of droplet impact. However, these soft hydrophobic have practical limitations, as the air layer is highly unstable and prone to collapse under impact, pressure, or in the presence of surface defects.

Inspired by the Nepenthes pitcher plant, Liquid Infused/Impregnated Surfaces (LIS)[22] and Slippery Liquid-Infused Porous Surfaces (SLIPs)[23] have been introduced as robust alternatives to superhydrophobic surfaces, where a lubricating oil film replaces air pockets. These surfaces offer ultra-low contact-angle hysteresis, mechanical self-healing, pressure stability, and resistance to fouling and icing, making them highly attractive for real-world applications.[22,24–29] With the successful biomimicry of these textured non-wetting surfaces, studies have explored their static and dynamic wettability characteristics, including droplet impact.[20,21,30] Droplet impact study on LIS by Choongyeop et al[31] investigated the dynamics of water droplet impact on oil-infused CuO nanostructured surfaces. They have observed that low-viscosity lubricants and droplet impact at higher Weber numbers more readily displace the lubricating film, triggering rim instabilities that limit the spreading radius and ultimately influence the rebound behavior. In our recent work,[26] silicone oils and hexadecane were used to prepare LIS. Both lubricants exhibit different intermolecular interactions with the textured silicon substrate, resulting in either Van der Waals or non-Van der Waals-impregnated states. These interactions governed the stability of the lubricant layer under droplet impact, therefore, the rebound behavior. The study highlights that the lubricant's physical and chemical properties critically determine LIS stability. A key limitation of LIS is that, even when the surface remains mechanically stable, the lubricant stored within the texture gradually depletes under dynamic conditions, resulting in a progressive loss of slipperiness. To address this issue, researchers have used SLIP surfaces[32] , which offer improved lubricant retention and enhanced long-term durability.

SLIP surfaces made of PDMS, when infused with a lubricant, have been studied for their wettability, particularly in applications aimed at preventing anti-icing and biofouling.[32] Researchers have examined the durability of LIS under various conditions. In one such study,[2]

Subsequent drop-impact experiments on SLIPs investigated the variations in oil-layer thickness, and the Weber number influences the spreading and rebound dynamics. Demonstrating that a sufficiently thick layer is necessary to prevent dewetting on the porous nanostructure and maintain uniform lubricant distribution. In our recent study,[33] smooth PDMS surfaces prepared via two distinct methods, lubricant coating and lubricant absorption, exhibited significantly different droplet impact outcomes. It was observed that merely coating a smooth surface with a lubricant did not meaningfully alter the impact dynamics compared to bare PDMS. However, when the lubricants were absorbed into the polymer matrix, the droplet impact dynamics could be systematically modified, with differences depending on the lubricant. This demonstrates that by selecting an appropriate lubricant, the impact behavior can be tailored to suit specific target applications. Similarly, several studies have highlighted that mixing oil with PDMS leads to different outcomes in droplet bouncing.[23]

However, no clear explanation or systematic study exists in the literature for lubricant-coated textured PDMS versus lubricant-absorbed textured PDMS surfaces. Studying this is important because PDMS is a porous matrix; the lubricant can readily absorb into the bulk, yet surface textures trap the oil, creating localized lubricant pools whose influence during droplet impact has not been reported. Furthermore, the combined effects of two different lubricant chemistries, solid fraction, and substrate softness on the static and dynamic wettability on porous soft surfaces remain unexplored in the existing literature. In this study, Textured PDMS surfaces were fabricated with square microposts measuring 10 × 10 × 10 µm (length × width × height) and post spacings of 5µm and 20µm. The influence of surface treatment and lubrication method on the wettability and droplet–surface interaction was then examined. Two lubricants, silicone oil (5cSt) and hexadecane, were introduced either by surface coating or by absorption into the textured PDMS. High-speed imaging was used to characterize droplet spreading, retraction, and rebound over a wide range of Weber numbers. The results explain how microtexture geometry, surface chemistry, and lubricant type collectively govern droplet impact dynamics. Overall, the findings demonstrate a straightforward approach to design SLIP surfaces with strong, durable water repellency, which could be useful for applications in droplet handling, self-cleaning, and microfluidic devices.

# 2. Experimental Section

## 2.1. Sample Preparation and Surface Characterization

Microtextured silicon surfaces with 10μm square posts, heights of 10μm (10 × 10 × 10 μm (length × width × height), and interpost spacings of 5μm and 20μm, were fabricated using a standard photolithographic process.[34] This microtextured silicon surface was used to fabricate microtextured PDMS surfaces using a soft lithography process.[35,36] The PDMS-textured surfaces were prepared using PDMS (Sylgard 184, Dow Corning, Wiesbaden, Germany). The commercially available liquid PDMS was mixed with a curing agent in a 10:1 ratio (PDMS to curing agent)[37]. The detailed fabrication process for textured PDMS from a textured silicon surface is described in the supporting information. These soft-textured PDMS surfaces were then functionalized using octadecyltrichlorosilane (OTS, Sigma-Aldrich) via a liquid phase deposition method[34] (explained in the supporting information). This treatment imparted non-polar, low-surface-energy characteristics to the surface, significantly affecting surface wettability, lubricant retention, and droplet interaction behavior.[19,38–42] From the contact angle measurements, OTS reports a free surface energy of 26mN/m.[43]

The choice of lubricant is essential for obtaining stable SLIPs. In addition, the lubricant should have a suitable affinity for binding to the surface. In this study, we selected two lubricants, one with a higher affinity for the OTS & PDMS surface and the other with a lower affinity. SLIPs formed with lubricants of higher surface affinity are referred to as Van der Waals SLIPs (VdW SLIPs), while those with lower affinity are termed non-Van der Waals SLIPs (nVdW SLIPs). The affinity of a lubricant for the OTS surface was assessed by measuring its equilibrium contact angle (Eq. CA) on the OTS-functionalized PDMS surface. For strong affinity, the Eq. CA must remain below 5° to confirm stable wetting of the lubricant on the surface.[44] We measured Eq.CA and contact angle hysteresis (CAH) of SO-5cSt and hexadecane on OTS-coated smooth PDMS surfaces in air and DI water using a Rame-Hart Model 500-U1 Advanced Goniometer. The lubricant was chosen such that it must be immiscible with water and have a similar viscosity to that of the impacting liquid (i.e., same order of magnitude). Thus, we chose two lubricants that fit our criteria: SO-5cSt and hexadecane.

Post spacing on the textured surface is another crucial parameter in our study. The stability of the SLIPs can be assessed based on the advancing and receding contact angles and the critical contact angle of the textured surface. Tables S2 and S3 in the supporting information provide detailed explanations of the measurement and calculation of the parameter described above. It is important to note that if the receding angle is greater than the critical contact angle, $\theta_{rec,os(a)} > \theta_c$, then the lubricant film will not spontaneously spread onto the textured surface,[44] i.e., the textured surface is unstable for that particular lubricant. However, when $\theta_{rec,os(a)} < \theta_c$, the lubricant film spreads onto the textured surface, and the impregnating liquid film remains stable.[44] Thus, to have a stable SLIPs configuration when coated with lubricant, we chose the 5µm and 20µm post spacing sample.

To evaluate the lubricant's affinity for the textured surface, the effective Hamaker constant was calculated using standard combining rules[45] (see the supporting information for details). This parameter represents the net Van der Waals interaction between the solid and the liquid and is commonly used to assess the stability of thin interfacial films. The sign of the effective Hamaker constant is particularly informative: a negative value indicates attractive interactions that favor the formation of a stable lubricant layer on the surface, whereas a positive value reflects repulsive interactions that hinder thin-film formation.[45] This analysis helps explain how different lubricants interact with the PDMS matrix and whether they can maintain a continuous film under dynamic conditions.

The samples were coated with lubricant by dipping them into a reservoir and then withdrawing them at a controlled speed (V) to maintain a ($Ca = \mu_o V / \gamma_{oa}$) capillary number of $10^{-5}$.[46] This method ensured a uniform lubricant coating without leaving excess oil on the surface. In this equation, $\mu_o$ is the dynamic viscosity and $\gamma_{oa}$ is the surface tension of the lubricant. By keeping the capillary number constant at $10^{-5}$, the withdrawal speed (V) was adjusted to maintain a constant lubricant thickness despite differences in viscosity. For both SO-5cSt and hexadecane, the withdrawal speed was selected based on the capillary number to ensure a consistent oil thickness across all experiments. In this manner, microtextured surfaces with varying post spacings were functionalized with OTS and subsequently coated with the two lubricants, as illustrated in Figure 1.

It has been observed that when textured PDMS surfaces are dip-coated with both lubricants, a stable configuration forms, i.e., the oil is retained between the microtextures. This was confirmed through microscopic imaging and theoretical calculations.[44] However, over time, the lubricant trapped between consecutive microtextures gradually gets absorbed into the PDMS matrix due to the material's inherent porosity.[24,47,33] This absorption leads to slight surface deformation, trapping the oil within the PDMS elastomer.[39,48] This makes the coated, textured PDMS surfaces transient liquid-infused surfaces. Four textured PDMS-OTS samples coated with lubricants were obtained. i.e. PDMS-OTS$_{(5\mu m)(SO\text{-}5cSt\text{-}coated)}$, PDMS-OTS$_{(20\mu m)(SO\text{-}5cSt\text{-}coated)}$, PDMS-OTS$_{(5\mu m)(Hexa\text{-}coated)}$, and PDMS-OTS$_{(20\mu m)(Hexa\text{-}coated)}$.

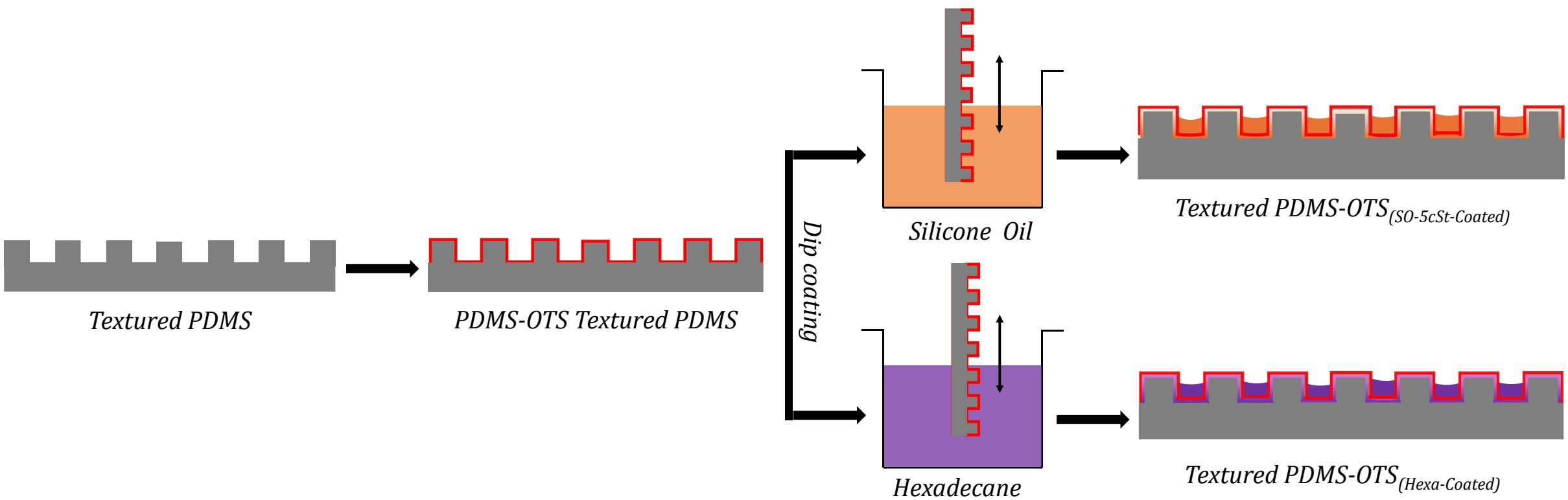

**Figure 1.** Schematic representation of the fabrication of textured PDMS-OTS when coated with SO-5cSt and hexadecane lubricant.

To investigate the effect of absorption on wettability and droplet impact, we added a sample set in which the textured PDMS surface was immersed in the lubricant. For the absorbed textured PDMS, the lubricant was allowed to absorb into the PDMS-OTS to study the effect of a bulk oil phase on the wettability and droplet impact. The textured PDMS surfaces were immersed in the oils (SO-5cst and hexadecane) for the absorption samples and left to soak for 24 hours.[36] Thus, four textured PDMS-OTS samples were absorbed with lubricants, as shown in Figure 2. i.e., PDMS-OTS$_{(5\mu m)(SO\text{-}5cSt\text{-}Absorb)}$, PDMS-OTS$_{(20\mu m)(SO\text{-}5cSt\text{-}Absorb)}$, PDMS-OTS$_{(5\mu m)(Hexa\text{-}Absorb)}$, and PDMS-OTS$_{(20\mu m)(Hexa\text{-}Absorb)}$. Before conducting the droplet-impact tests, the oil-absorbed samples were dip-coated to remove excess lubricant and ensure a uniform oil layer on the surface.[46]

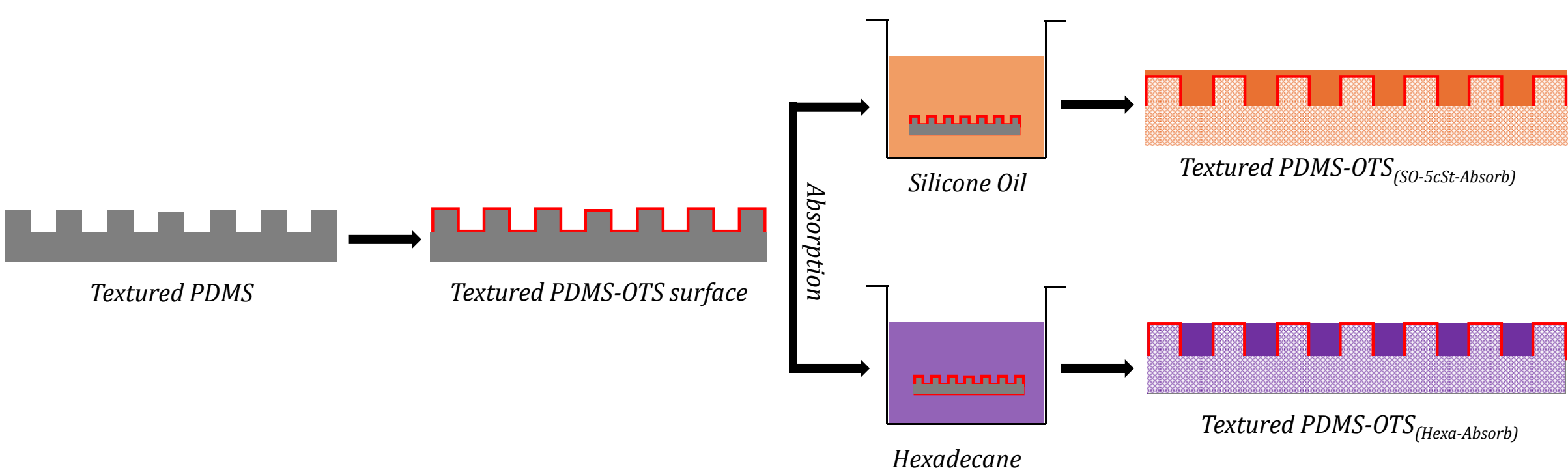

**Figure 2.** Schematic representation of the fabrication of textured PDMS-OTS when absorbed with SO-5cSt and hexadecane lubricant.

## 2.2. Droplet Impact Experiments

Droplet impact experiments were carried out by placing the samples on a flat metal sliding stage. Droplets with a diameter of 2.8 mm were produced from the tip of a Teflon-coated needle connected to a syringe pump, which was operated at an infusion rate of 1 ml/hr using a Harvard Apparatus syringe pump. The impact velocity ($Vi$) of the droplets was controlled by varying the fall height between 4 and 70 cm, giving velocities from 0.88 to 3.70 m/s. Initial tests showed distinct droplet behaviors within this velocity range. Based on observations, four Weber numbers ($We$) were selected: 28, 63, 127, and 247, spanning a broad range of droplet impact outcomes, including deposition, partial rebound, full rebound, prompt splash, and prompt splash with full rebound. The Weber number, defined as $We = (\rho D Vi^2/\sigma)$, represents the ratio of inertial to surface tension forces, where $\sigma$ is the surface tension, $\rho$ is the water density, and $D$ is the droplet diameter. Droplet impact dynamics were recorded from the side using a Phantom VEO 410 high-speed camera at 1280 × 720 resolution and 5000 frames per second. A high-intensity light source was positioned behind the substrate, ensuring that the light, the substrate, and the camera were aligned along the same optical axis, as illustrated in Figure 3. Video analysis was performed with MATLAB, while ImageJ was used to extract data from images representing different stages of droplet impact.[49] In total, 250 videos were captured and analyzed to improve the accuracy of droplet impact measurements.

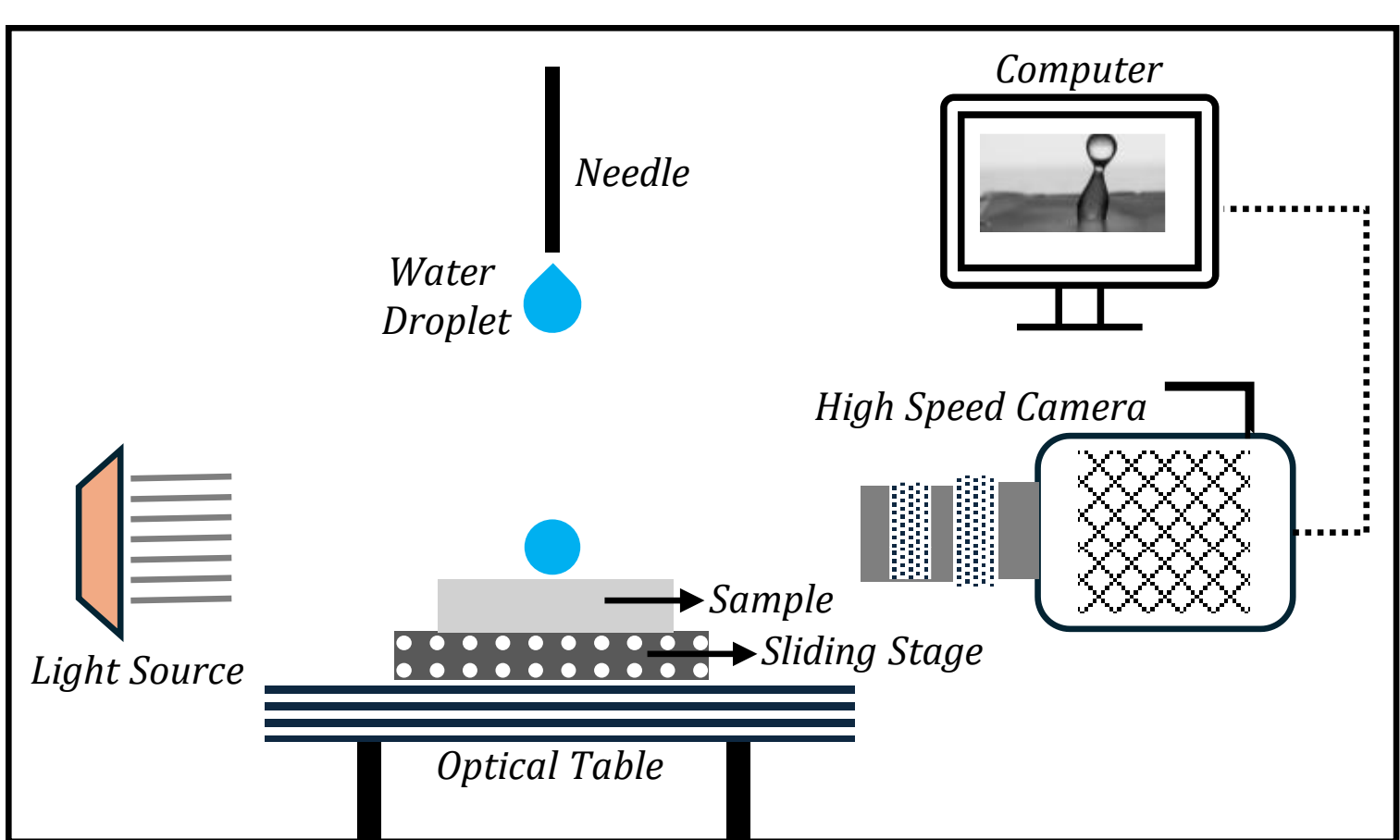

**Figure 3.** Schematic diagram of droplet impact setup.

# 3. Results and Discussions

## 3.1. Stability and Surface Wettability

The stability of the oil infused into the textured PDMS surface can be explained within a thermodynamic framework by considering interfacial energies at distinct interfaces. There are 12 possible configurations in a three-phase system in which oil impregnation occurs.[44] This thermodynamic framework allows us to predict which of the 12 states will be stable for a droplet, an oil, and a substrate. Considering contact angle measurements, both Hexadecane and SO-5cSt will exhibit stable configurations in air and water environments for post spacings of 5μm and 20μm. In a four-system involving the droplet, oil, solid substrate, and surrounding medium, we identified four stable configurations (as illustrated in Table S6 and Figure S4(a and b)): two beneath the droplet and two along its sides. Specifically, both hexadecane and SO-5cSt oils exhibit stable configurations in air and water environments across post spacings of 5 and 20μm. The evaluation of the effective Hamaker constant for the substrate–oil interaction shows that silicone oil possesses a negative value ($-0.1801 \times 10^{-20}$ J) on OTS-coated surfaces, which satisfies the criterion for forming a stable thin film. In contrast, hexadecane exhibits a positive effective Hamaker constant ($0.66 \times 10^{-20}$ J) (Shown in Table S7), indicating that thin film formation is not favored. These theoretical predictions are consistent with the experimental observations.

Figure 4 illustrates the measured contact angle hysteresis (CAH) for PDMS-OTS surfaces under different configurations. Figure 4. (a) shows CAH values for textured PDMS-OTS substrates with post spacings of 5μm and 20μm. An increase in contact angle hysteresis (CAH) was observed as the post spacing increased. This behavior is attributed to enhanced contact line pinning at larger spacings, where the collapse of air pockets allows greater interaction between the liquid and the underlying substrate. At a spacing of 5 μm, an air cushion persists beneath the droplet (Cassie–Baxter state), minimizing pinning and resulting in low CAH. As the spacing increases to 20 μm, the surface transitions to a Wenzel state, in which water infiltrates the texture, replacing the trapped air. This leads to increased distortion of the contact line during both advancing and receding motion, thereby producing higher hysteresis.

Figure 4(b) shows the CAH for the same surfaces coated with two lubricants, SO-5cSt and hexadecane. For a 5μm post spacing, there is little significant change in the CAH when compared with uncoated textured PDMS. This is because the air layer beneath the droplet is now replaced by lubricating oil, and the water droplet doesn't come in contact with the lubricant. In 20 μm textured surface lubricant-coated samples, infusion significantly reduces CAH. This occurs because the infused oil fills the post spacing, reducing contact line pinning and thereby lowering CAH compared to the uncoated 20μm PDMS surface.

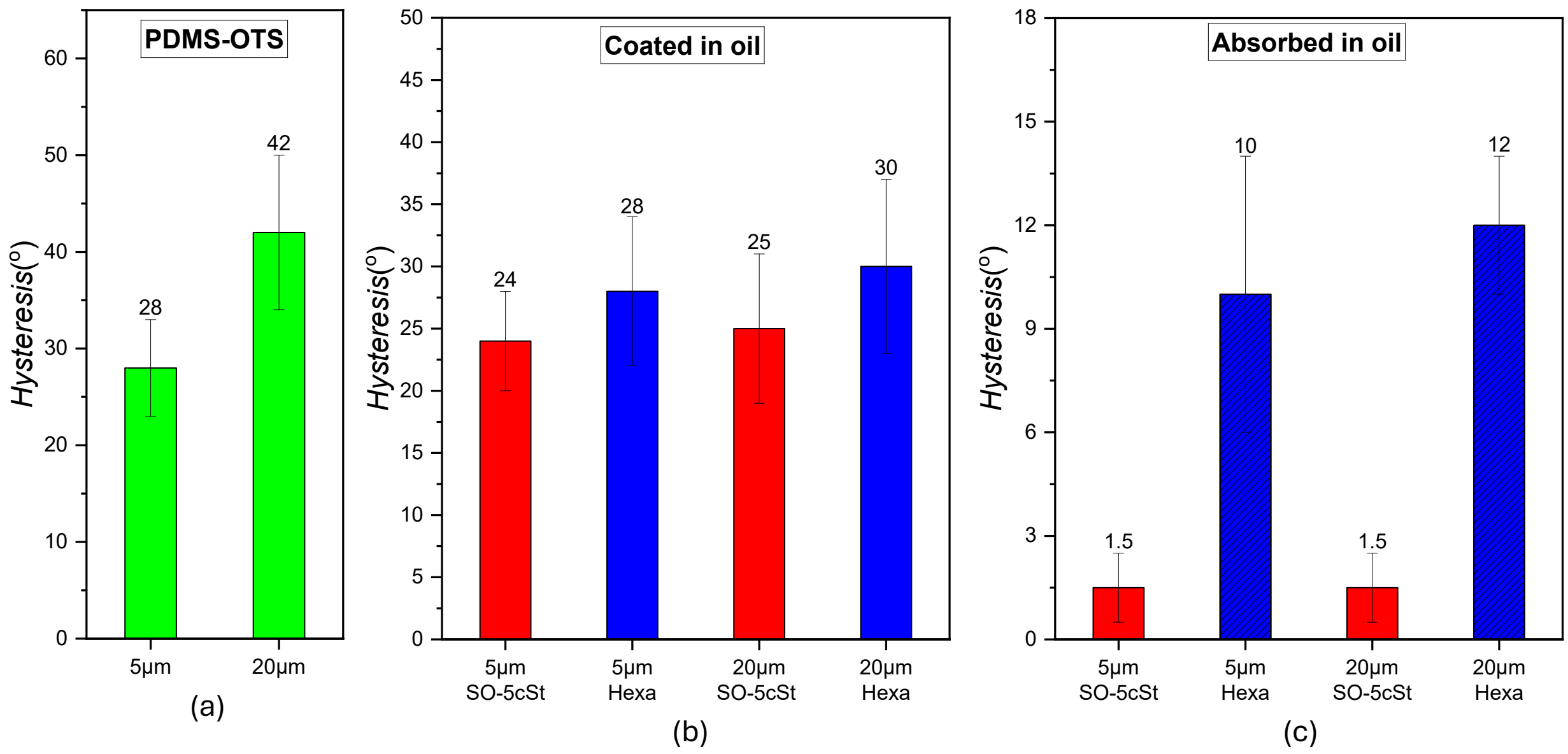

**Figure 4.** Summary of contact angle hysteresis data for 5μm and 20μm post spacing for (a) PDMS-OTS samples, (b) PDMS-OTS samples coated with different lubricants, and (c) PDMS-OTS samples absorbed with different lubricants. Error bars indicate the standard deviation from ten independent measurements.

Silicone oil-coated surfaces with 5 and 20μm post spacing exhibit only a very slight decrease in CAH compared to hexadecane-coated samples. This behavior arises from the chemical similarity between silicone oil and PDMS, which promotes a stronger affinity. However, this affinity does not significantly influence the wettability of the textured PDMS surface when coated with different lubricants. This is because of the porous nature of PDMS, where the lubricant on the top of the posts is rapidly absorbed into the matrix, exposing the OTS-functionalized tops directly to air. In contrast, the space between the posts acts as micro-reservoirs that retain oil for longer. Although the oil in these gaps also gradually absorbs into the PDMS, the confined geometry delays this process, leaving a thin lubricant film within the valleys as seen in Figure 5, which illustrates oil distribution on the textured PDMS-OTS surfaces after coating. This uneven absorption results in a hybrid wetting state, where the droplet experiences a mix of lubricated and exposed regions during impact or spreading. This diminishes the overall lubricating efficiency and results in larger CAH. To summarise the CAH trends observed for the coated samples: PDMS-OTS$_{(20\mu m)}$ > PDMS-OTS$_{(5\mu m)}$ > PDMS-OTS$_{(20\mu m)(Hexa\text{-}coated)}$ > PDMS-OTS$_{(5\mu m)(Hexa\text{-}coated)}$ > PDMS-OTS$_{(5\mu m)(SO\text{-}5cSt\text{-}coated)}$ > PDMS-OTS$_{(20\mu m)(SO\text{-}5cSt\text{-}coated)}$.

The CAH measurements for textured PDMS-OTS samples absorbed with lubricant revealed interesting trends, as shown in Figure 4(c). The SO-5cSt absorbed PDMS-OTS samples for 5µm and 20µm post-spacing; both, i.e., (PDMS-OTS$_{(5\mu m)(SO\text{-}5cSt\text{-}Absorb)}$ and (PDMS-OTS$_{(20\mu m)(SO\text{-}5cSt\text{-}Absorb)}$) surfaces exhibited significantly lower CAH values. This behavior can be attributed to the complete absorption of low-viscosity oil into the substrate and to its trapping between the posts due to capillary forces. This forms a thin lubricant film at the surface (Figure 6(a)). This continuous oil film reduces the solid-water interfacial interaction, resulting in lower CAH. In contrast, the CAH values for textured PDMS-OTS samples for 5µm and 20µm post-spacing absorbed with hexadecane, i.e., for PDMS-OTS$_{(5\mu m)(Hexa\text{-}Absorb)}$, PDMS-OTS$_{(20\mu m)(Hexa\text{-}Absorb)}$ are significantly higher than those observed for SO-5cSt absorbed samples. This increased CAH can be attributed to the absence of a stable thin lubricant film at the surface.[29] However, the oil present between the posts gives sufficient lubrication to the water droplet. Thus, the CAH of textured PDMS-OTS absorbed in hexadecane has a lower CAH when compared with the textured PDMS-OTS sample coated with hexadecane (see Figure 4(b & c)). To summarise the CAH trends observed for the lubricant absorbed samples: PDMS-OTS$_{(20\mu m)}$ > PDMS-OTS$_{(5\mu m)}$ > PDMS-OTS$_{(5\mu m)(Hexa\text{-}Absorb)}$ > PDMS-OTS$_{(20\mu m)(Hexa\text{-}Absorb)}$, PDMS-OTS$_{(5\mu m)(SO\text{-}5cSt\text{-}Absorb)}$ > PDMS-OTS$_{(20\mu m)(SO\text{-}5cSt\text{-}Absorb)}$.

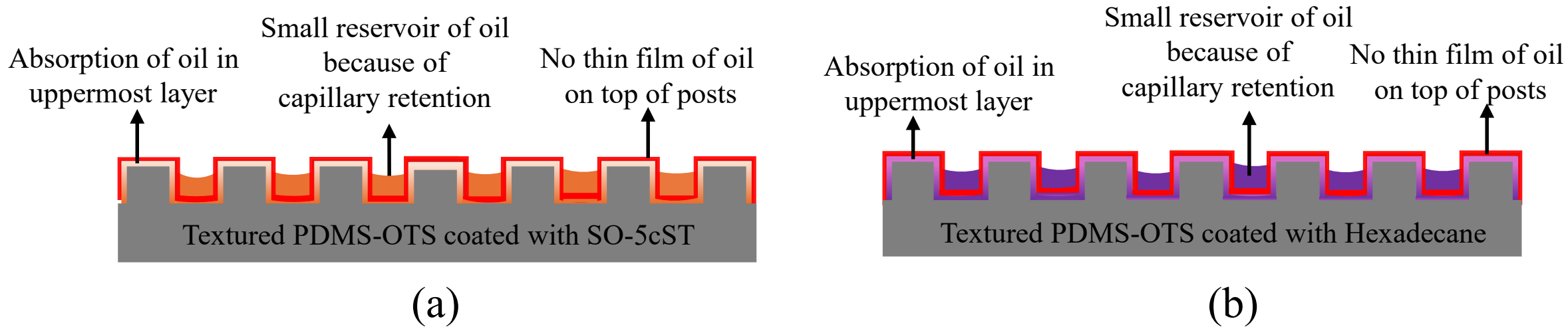

**Figure 5.** Schematic representation of the lubricant on coated OTS-functionalized textured PDMS surfaces: (a) SO-5 cSt (orange) and (b) Hexadecane (purple). The red line indicates the OTS-functionalized layer on the textures.

Figure 6 illustrates lubricant distribution after 24 hours of absorption into the textured PDMS-OTS surface. During this period, oil is absorbed into the porous PDMS matrix until it reaches its saturation limit. For the SO-5cSt lubricant (left), the oil can penetrate deeply and uniformly into the entire PDMS matrix, including the textured region and the top surface. This results in the formation of a stable, continuous thin lubricating film, known as a SLIPs (Slippery Liquid-Infused Porous Surface), supported by the oil's favorable interaction with the PDMS-OTS. Consequently, SO-5cSt is present both in the texture grooves and on the tops of the posts(Figure 6(a)), creating a uniformly lubricated surface that remains stable in both air and water environments. In contrast, for hexadecane (Figure 6(b)), although bulk absorption into the matrix occurs over 24 hours, the lack of a strong interaction with PDMS-OTS inhibits thin film formation on the top surface. Instead, the oil remains primarily within the grooves between posts, forming a stable but localized reservoir of lubricant. The top surface of the posts (PDMS-OTS) remains exposed to air due to insufficient film formation. This results in a hybrid wetting state during droplet interaction, where the droplet partially contacts the exposed solid (i.e., PDMS-OTS) and partially interacts with the lubricated grooves. The theoretical calculation, wettability measurements, and CAH analysis supported such behavior.

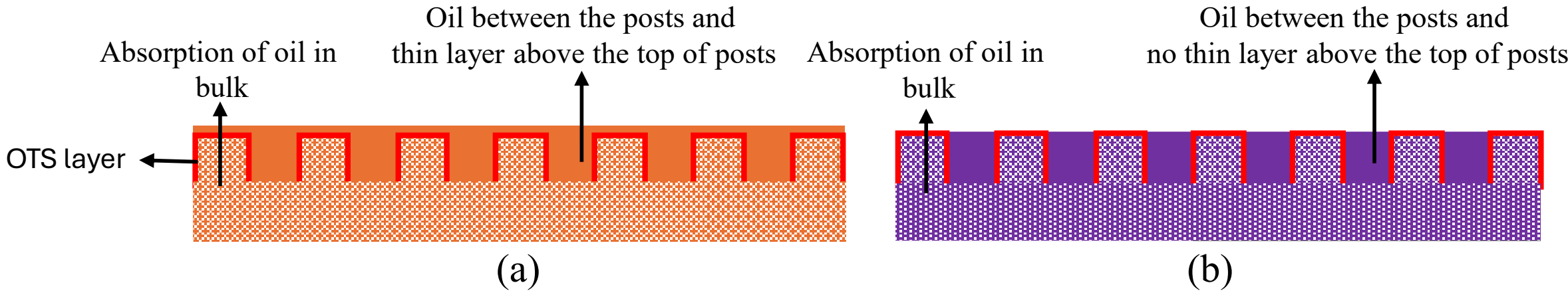

**Figure 6.** Schematic representation of the lubricant absorption into OTS-functionalized textured PDMS (a) SO-5 cSt (orange) and (b) Hexadecane (purple). The red line indicates the OTS functionalized top layer, while the shaded region represents oil absorbed into the PDMS matrix.

Before analyzing droplet impact dynamics on textured PDMS-OTS surfaces coated or absorbed with SO-5cSt and hexadecane, it is essential to understand the baseline behavior first. This involves studying droplet impact on textured PDMS-OTS surfaces without any lubricant. This comparison provides critical insight into the role of lubricant in altering wetting, spreading, and rebound behavior. Figure 7(a and b) illustrates droplet impact on OTS functionalized PDMS surfaces with 5μm and 20μm post spacing at different Weber numbers, respectively. For a 5μm spacing, a complete rebound is observed at lower Weber numbers, as the droplet cannot penetrate the air pockets, due to the high capillary pressure, and remains in the Cassie-Baxter state. At higher Weber numbers, partial rebound occurs as water penetrates the texture. In contrast, droplets exhibit complete deposition on the 20μm surface even at lower Weber numbers. The larger post spacing reduces the capillary barrier, allowing the droplet to enter the texture and transition toward a Wenzel state. This leads to greater energy dissipation and suppresses rebound. As a result, kinetic energy is insufficient to support rebound. These trends underscore the crucial role of texture spacing in determining the wetting state and its influence on the outcome. The theoretical explanation for this is provided in the supporting information. These results are consistent with previous research on droplet impact on textured surfaces.[19–21,26,50,51]

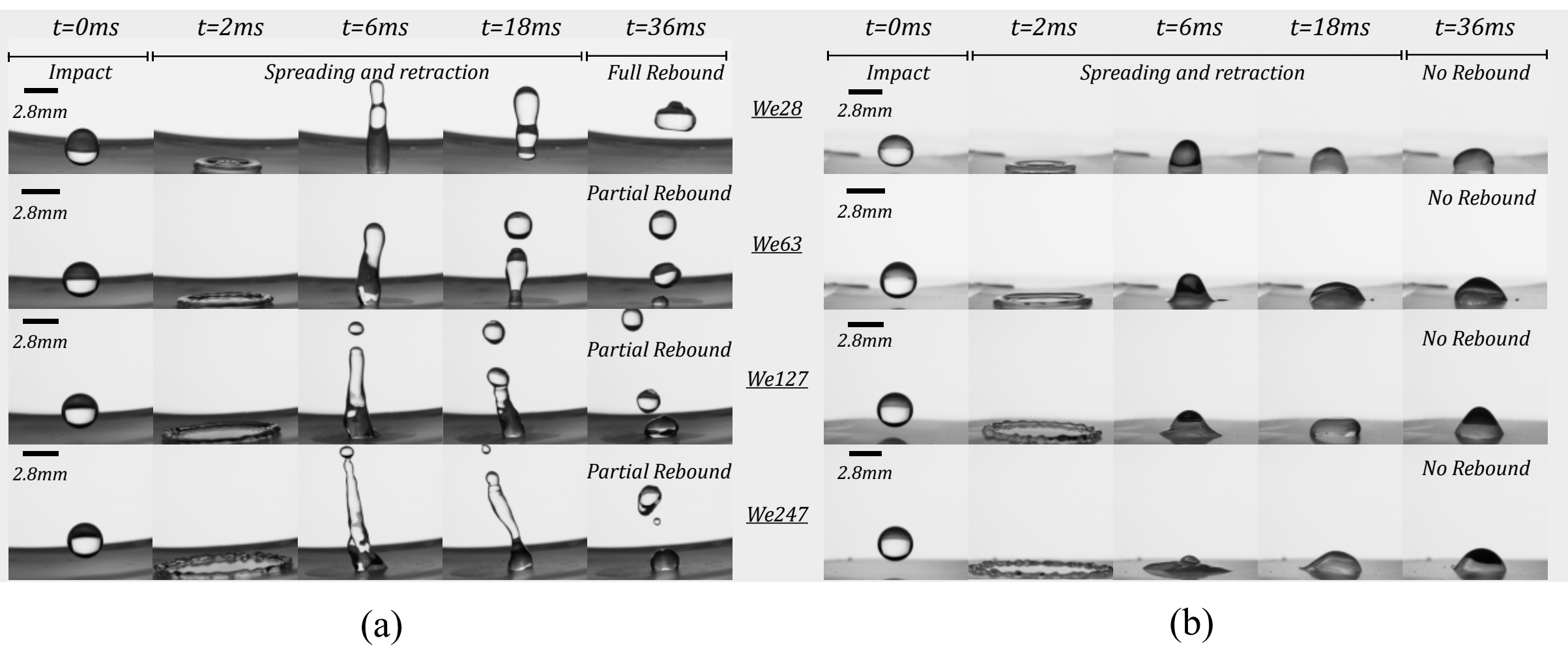

(a) (b)

**Figure 7.** Time-resolved images showing droplet impact dynamics on PDMS-OTS surfaces with (a) 5 µm and (b) 20 µm post spacing across varying Weber numbers.

### 3.2. Effect of Lubricant Coating

In our previous study,[26] Solid textured silicone surfaces were coated with silicone oil and hexadecane to create Van der Waals and non-Van der Waals liquid-infused surfaces.[26] When applying the same approach to soft PDMS textured surfaces, it was initially expected that the lubricant would remain trapped between adjacent micro-posts, forming stable reservoirs similar to those in solid/conventional liquid-infused surfaces. However, observations with an optical microscope, contact angle measurements, and droplet impact experiments showed that the oil was retained between the posts for only a limited period. This behavior is attributed to the intrinsic porosity of PDMS, which enables the lubricant to gradually absorb into its upper layer. Such absorption has been reported previously and is consistent with PDMS's well-recognized porous characteristics.[36,38] This reduces the amount of oil between the posts, as shown in Figure 5. As a result, the top surface of the microstructures remains predominantly exposed to air, and a very small amount of oil is present between the textures, leading to a heterogeneous wetting regime in which both solid-air and liquid-oil interactions coexist. Which eventually leads to different rebound outcomes. The results of the droplet impact experiments on textured PDMS surfaces functionalized with OTS and subsequently coated with different lubricants for post spacings of 5 and 20μm are illustrated in Figures 8 and 9, respectively.

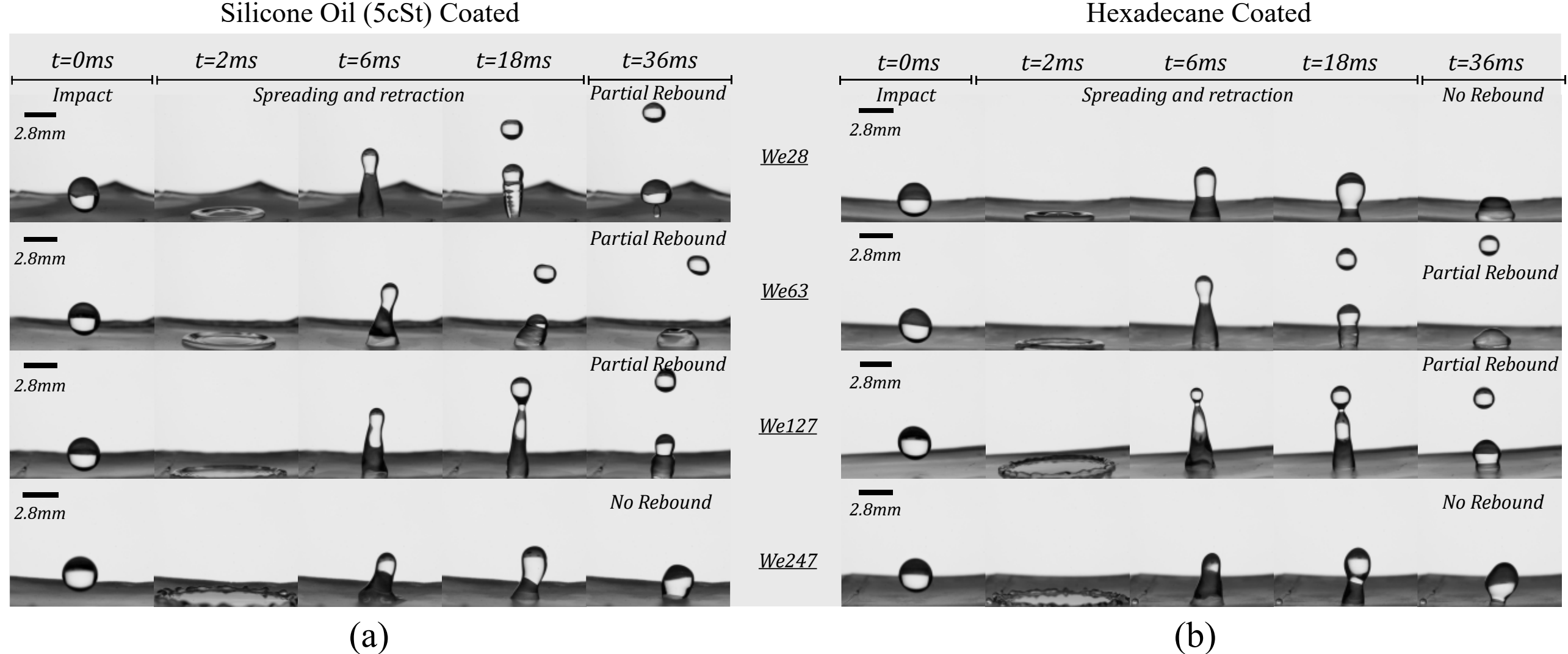

**Figure 8.** Time-resolved images showing the droplet impact on textured PDMS-OTS surfaces with a post spacing of 5μm when coated with different oils: (a) silicone oil (5cSt) and (b) hexadecane, shown at various Weber numbers.

Figure 8 illustrates the droplet impact dynamics on the post spacing of 5 μm for the range of Weber numbers. At a lower Weber number (*We* = 28), coated with lubricants SO-5cSt and hexdecane (Figure 8 (a & b), distinct differences in droplet behavior are observed between uncoated PDMS-OTS textured surfaces (Figure 7(a & b) (*We*=28)). During impact, due to heterogeneous wetting, oil reservoirs within the gaps provide localized viscous damping, while the exposed micro post tops hinder spreading. This asymmetric surface condition introduces drag and resistance to the droplet's lateral motion during the spreading and retraction phases, ultimately suppressing the recoil and resulting in partial rebound. In contrast, on the uncoated PDMS-OTS surface for 5μm ( Figure 7 (a & b)), characterized by a trapped air cushion between the microstructures. This allows the droplet to partially infiltrate the textured structure.

At intermediate *We* = 63 and 127, for both lubricants shown in Figure 8 (a & b), the bouncing behavior on lubricant-coated surfaces becomes qualitatively similar to that observed on uncoated PDMS-OTS surfaces for *We* = 63 and 127 (see Figure 7(a & b)). However, the degree of rebound, characterized by reduced rebound height and spreading diameter, remains lower. This behavior can be attributed to enhanced viscous dissipation from the droplet's contact with the residual oil in the textured gaps. The lubricant provides additional resistance to the recoiling lamellae, thus damping the rebound. At higher Weber numbers (*We* = 247), 5 μm-coated with the SO-5cSt lubricant (Figure 8(a), partially enter the texture and push some oil out of the gaps. This results in additional energy loss due to viscous resistance. As a result, the droplet sticks and cannot slide back, showing no rebound. Similarly, for *We* = 247 in hexadecane-coated PDMS-OTS surfaces, the bouncing behavior is not different, as hexadecane exhibits a comparatively weaker affinity for the PDMS-OTS matrix due to lower van der Waals interaction. When a droplet impacts the hexadecane-coated surface, the kinetic energy drives the water into the texture, rapidly displacing the weakly bound hexadecane. As the droplet penetrates and displaces the oil, it directly interacts with the solid PDMS-OTS substrate. This increases adhesion and energy dissipation, causing the droplet to complete deposition rather than partial rebound, as observed on an uncoated PDMS-OTS 5μm textured surface (see Figure 7(b) *We* = 247). Thus, the PDMS-OTS 5μm textured surface coated with hexadecane and SO-5cSt behaves like an unlubricated hydrophobic substrate under high Weber number impacts.

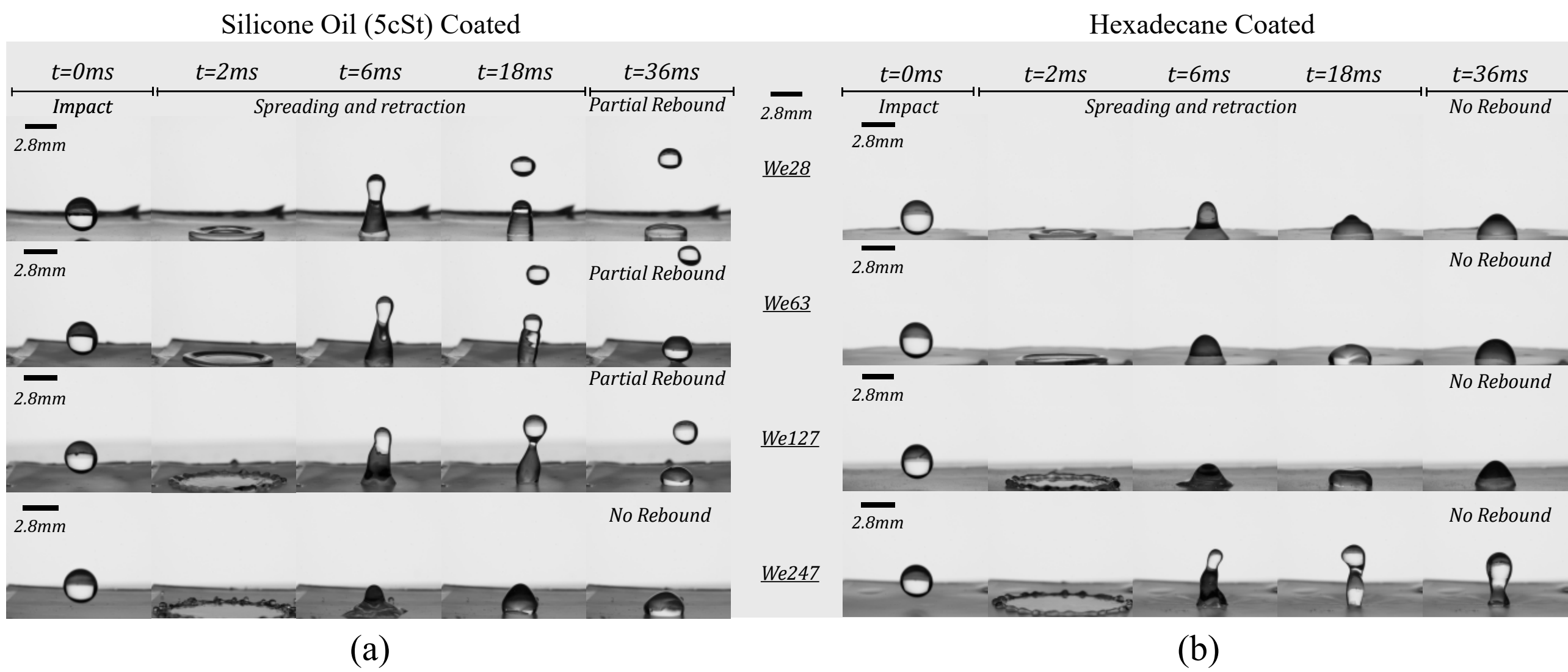

**Figure 9**. Time-resolved images showing the droplet impact on textured PDMS-OTS surfaces with a post spacing of 20μm when coated with different oils: (a) silicone oil (5cSt) and (b) hexadecane, shown at various Weber numbers.

Figure 9 illustrates the droplet impact behavior on 20μm textured PDMS-OTS surfaces coated with SO-5cSt and hexadecane lubricants. For the SO-5cSt-coated surface, the partial rebound was consistently observed across the entire range of Weber numbers, except at higher Weber numbers. This behavior can be attributed to the droplet penetrating the textured surface during impact and displacing the excess lubricant stored in the texture gaps. However, a residual oil film remains at the interface, preventing direct contact between the water droplet and the PDMS-OTS substrate, thereby enabling partial rebound. However, at higher Weber numbers, this residual oil layer is also removed, resulting in complete adhesion of the water droplet to PDMS-OTS and no rebound.

The droplet impact on the hexadecane-coated 20μm textured PDMS-OTS surface resulted in complete deposition. During impact, the droplet readily displaced the loosely bound hexadecane from the texture gaps, allowing direct water contact with the PDMS-OTS surface. Since the inherent stickiness of PDMS-OTS dominates in the absence of an oil barrier layer, the droplet adheres and does not rebound, similar to its behavior on the uncoated PDMS-OTS surface (see Figure 7(b) for all Weber numbers). These observations underscore the crucial influence of both lubricant type and micro-post spacing on droplet impact behavior over textured, oil-absorbing PDMS surfaces. The lubricant's ability to form a stable film and its retention within the textured matrix directly affect spreading, rebound, and deposition outcomes. The distinct responses observed with SO-5cSt and hexadecane highlight how variations in oil retention and film stability influence overall impact dynamics, driven by interactions between the lubricant and the porous PDMS substrate.

### 3.3. Effect of Lubricant Absorption

To investigate the effect of droplet impact dynamics on lubricant-absorbed PDMS-OTS surfaces, samples were prepared by allowing the substrates to uptake SO-5cSt silicone oil and hexadecane, respectively, following the protocol described in the experimental section. For textured PDMS-OTS surfaces with 5μm and 20μm particles absorbed with hexadecane (See Figures 10(b) and 11(b)), different impact outcomes were observed, including complete rebound, partial rebound, and no rebound, depending on the Weber number and post spacing. At a post spacing of 5μm (Figure 10 (b)), where the solid fraction is high, and no stable thin film forms over the square micro-posts (for hexadecane), droplets exhibited partial rebound at lower and intermediate Weber numbers (i.e., $We \approx 28$ & $We \approx 63$). This indicates that the lubricating hexadecane trapped between two consecutive posts enables the droplet to glide over the surface, while the available kinetic energy is insufficient to penetrate the oil retained within the surface texture. Additionally, the inherent adhesive forces of the PDMS–OTS surface at the exposed tops of the posts resist droplet motion, thereby contributing to the partial rebound of the impacting water droplet. However, at a higher Weber number ($We \approx 247$), the droplet sticks to the surface entirely. This non-rebounding behavior can be attributed to the weak affinity between hexadecane and the PDMS-OTS substrate. During high-speed impacts, this weak interaction allows the water droplet to displace hexadecane from the textured valleys and to penetrate the gaps between posts, thereby establishing direct contact with the solid substrate. This solid-water contact enhances droplet pinning and adhesion, effectively suppressing rebound and resulting in complete deposition. Unlike SO-5cSt, which has stronger van der Waals interactions with PDMS-OTS, hexadecane does not form a robust lubricant film.

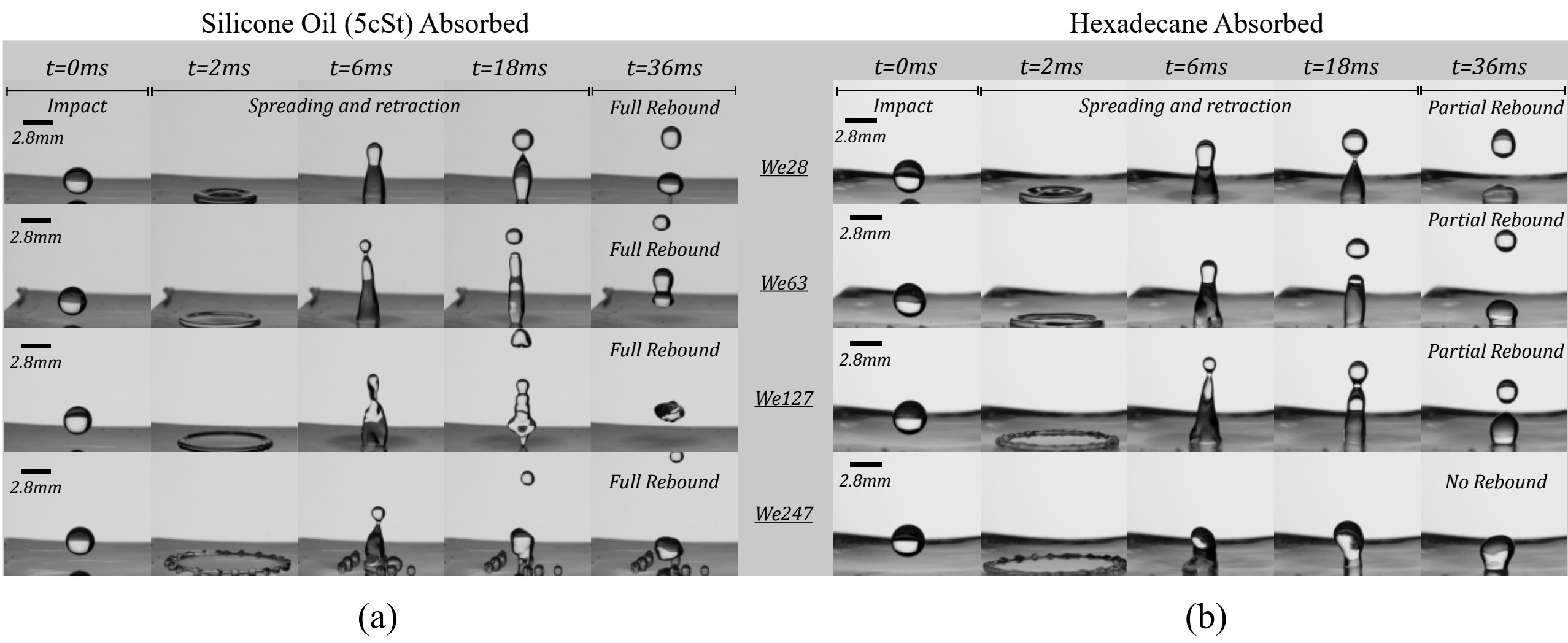

**Figure 10.** Time-resolved images showing the droplet impact on textured PDMS-OTS surfaces with a post spacing of 5μm when absorbed with different oils: (a) silicone oil (5cSt) and (b) hexadecane, shown at various Weber numbers.

Figure 11(b) shows a distinct trend in bouncing behaviour, ranging from partial to complete rebound, and no rebound was observed for the 20μm post-spacing when absorbed in the hexadecane. At low Weber numbers, droplets exhibited partial rebound due to insufficient kinetic energy to overcome surface adhesion, see Figure 11 (b) $We \approx 28$). The complete rebound was observed in the intermediate Weber number range ($We \approx$ 63-127), Figure 11 (b). This is primarily due to the lower solid fraction of the 20μm post-spacing texture, which reduces the contact area between the PDMS-OTS and the water droplet and facilitates easier droplet movement. The oil, absorbed primarily into the porous PDMS matrix, also provides minimal resistance at the top surface, allowing efficient recoil. However, droplets impacting hexadecane-absorbed surfaces exhibited complete deposition (Figure 11(b), $We \approx 247$). This is due to hexadecane's weak affinity for the PDMS-OTS surface, which allows the lubricant to be displaced by the water droplet during impact, leading to direct contact between the textured substrate and water. This enhanced pinning and adhesion at high-impact velocities suppress rebound, leading to complete droplet deposition.

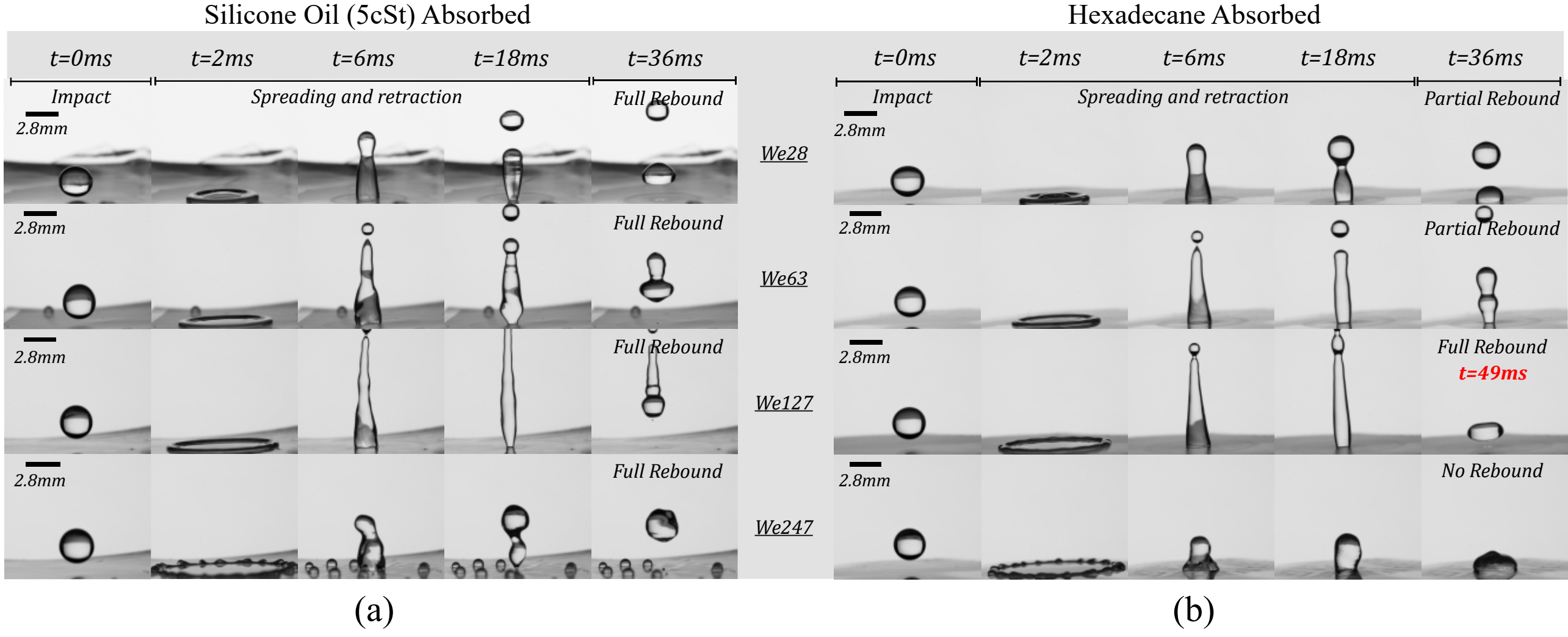

**Figure 11.** Time-resolved images showing the droplet impact on textured PDMS-OTS surfaces with a post spacing of 20μm when absorbed with different oils: (a) silicone oil (5cSt) and (b) hexadecane, shown at various Weber numbers.

Whereas the water droplet impact on the SO-5cSt absorbed textured PDMS-OTS surface revealed complete rebound across the full range of Weber numbers tested for both post spacings of 5 μm and 20 μm, as shown in Figures 10(a) and 11(a). This rebound behavior indicates that the textured PDMS-OTS absorbed with SO-5cSt oil forms a stable lubricating layer that spreads evenly over the top of the square post and within the microstructured gaps (i.e., between posts), enhancing surface uniformity and stability. The layer of oil on the top of the square post is sufficiently thick to prevent exposure of bare PDMS-OTS, yet thin enough to avoid significant viscous drag on the impacting droplet. As a result, the surface provides a low-adhesion, low-friction interface that facilitates efficient droplet rebound by reducing both contact-line pinning and viscous dissipation. At $We \approx 127$, the droplet spreads upon impact, then retracts, and ultimately rebounds from the surface without splashing. At higher Weber numbers, particularly around $We \approx 230$, an additional dynamic behavior was observed during the droplet spreading and retraction phase for 5 μm and 20 μm post-spacing samples absorbed with SO-5cSt (see Figure 10(a) & 11 (a) for $We \approx 247$). Two distinct instability mechanisms coexist during droplet impact.

Figure 12 shows droplet impact at ($We \approx 240$) on 5μm post spacing when absorbed with silicone oil (5cSt), the droplet no longer spreads smoothly but undergoes prompt splashing (shown in Figure 12(a)). The characteristics and intensity of this prompt splash are governed by the thermophysical properties of both the lubricant and the impacting droplet, including viscosity, density, and surface tension,[52] as well as the impact dynamics, such as the impact velocity. In addition, the surface roughness and texture geometry play an important role in promoting or suppressing splashing by influencing the stability of the spreading lamella.[53] During this regime, the expanding lamella becomes unstable and ejects small droplets from finger-like protrusions that form along the rim of the spreading liquid sheet (shown in Figure 12(b)). As the droplet spreads radially, the rim of the lamella experiences perturbations that amplify and evolve into azimuthal undulations. The formation of these fingers has often been interpreted using the Rayleigh-Taylor (RT) instability framework,[54] which describes the instability that occurs when a denser fluid is accelerated by a lighter one. In the context of droplet impact, the decelerating lamella creates an effective air acceleration directed toward the substrate, promoting instability at the liquid–air interface along the rim. In this scenario, inertial forces amplify perturbations, while surface tension stabilizes the interface and suppresses small-wavelength disturbances (shown in Figure 12(c)).[55,56]

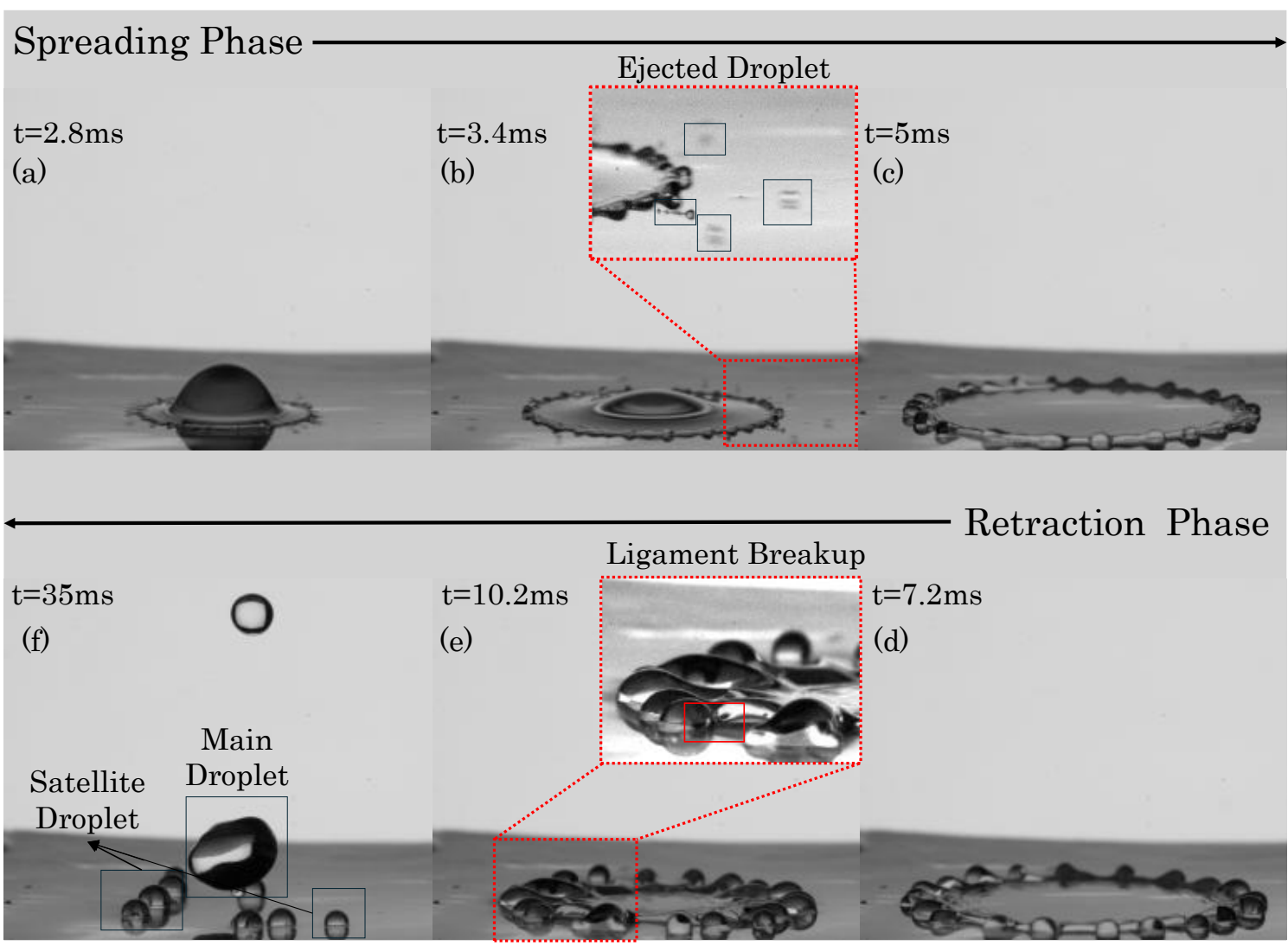

**Figure 12.** Instability at the spreading and retraction phases during droplet impact on a textured PDMS–OTS surface with a post spacing of 5 μm, absorbed with silicone oil (5cSt) at a Weber number of 247. (See supporting video 1)

Following the spreading phase, the droplet enters a rapid retraction phase driven by the extremely low contact angle hysteresis provided by the thin oil layer on the PDMS substrate. The reduced pinning at the contact line allows the liquid to retract with high velocity. As the liquid mass moves inward, fluid accumulates at the receding three-phase contact line, where both the interface curvature and local velocity gradients reach their maximum. This rapid inward motion generates a hydrodynamically unstable rim along the droplet perimeter (shown in Figure 12(d)). The rim becomes susceptible to perturbations, leading to the formation of slender, thread-like liquid filaments extending outward from the edge of the retracting droplet. These filaments are inherently unstable and rapidly undergo the Rayleigh-Plateau instability,[57,58] a capillary-driven mechanism in which a cylindrical liquid column breaks into smaller droplets in order to minimize its surface energy. As a result, the liquid threads rupture and pinch off into a series of small daughter droplets (shown in Figure 12(e)). These droplets appear momentarily as a ring of daughter droplets surrounding the central liquid mass, after which the remaining fluid recoils inward and forms a vertical jet at the droplet center (shown in Figure 12(e)).

Despite this transient rim instability for 5µm and 20µm post-spacing samples absorbed with SO-5cSt (see Figure 10(a) & 11 (a) for *We* = 247). The droplet retains sufficient momentum and coherence to completely recoil from the surface at high-impact velocities in both post-spacing cases. The ability of the surface to support such rebound, even after rim fragmentation, highlights the strong compatibility and interaction between the SO-5cSt lubricant and the OTS-functionalized PDMS. The oil is not visibly displaced or removed during impact, suggesting that the lubricant layer remains stable and intact throughout the process. The oil absorbed into the bulk of the PDMS helps maintain a continuous, slippery interface that resists water adhesion even under high-kinetic-energy conditions. These results collectively indicate a strong intermolecular affinity between SO-5cSt, OTS, and the PDMS substrate, which contributes to the stable retention of the lubricant and enhances the surface's anti-wetting performance.

### 3.4. Influence of Weber Number on Drop Impact Dynamics

The Weber number, which summarises the influence of impact velocity relative to surface tension, is crucial in determining droplet impact dynamics. A series of experiments was conducted across a range of Weber numbers to investigate its effect, with a focus on how it governs the spreading and retraction behavior of impacting droplets. Figure 8, Figure 9, Figure 10, and Figure 11 present sequential time-lapse images that capture the stages of droplet deformation and rebound on surfaces with 5µm and 20µm post spacings, each functionalized with OTS and subsequently coated with either SO-5cSt or hexadecane. One of the most common methods for analyzing the influence of the Weber number is to monitor the droplet's maximum spreading diameter. As depicted in Figures 13 and 14, this diameter increases consistently with increasing Weber number across all surface types studied.

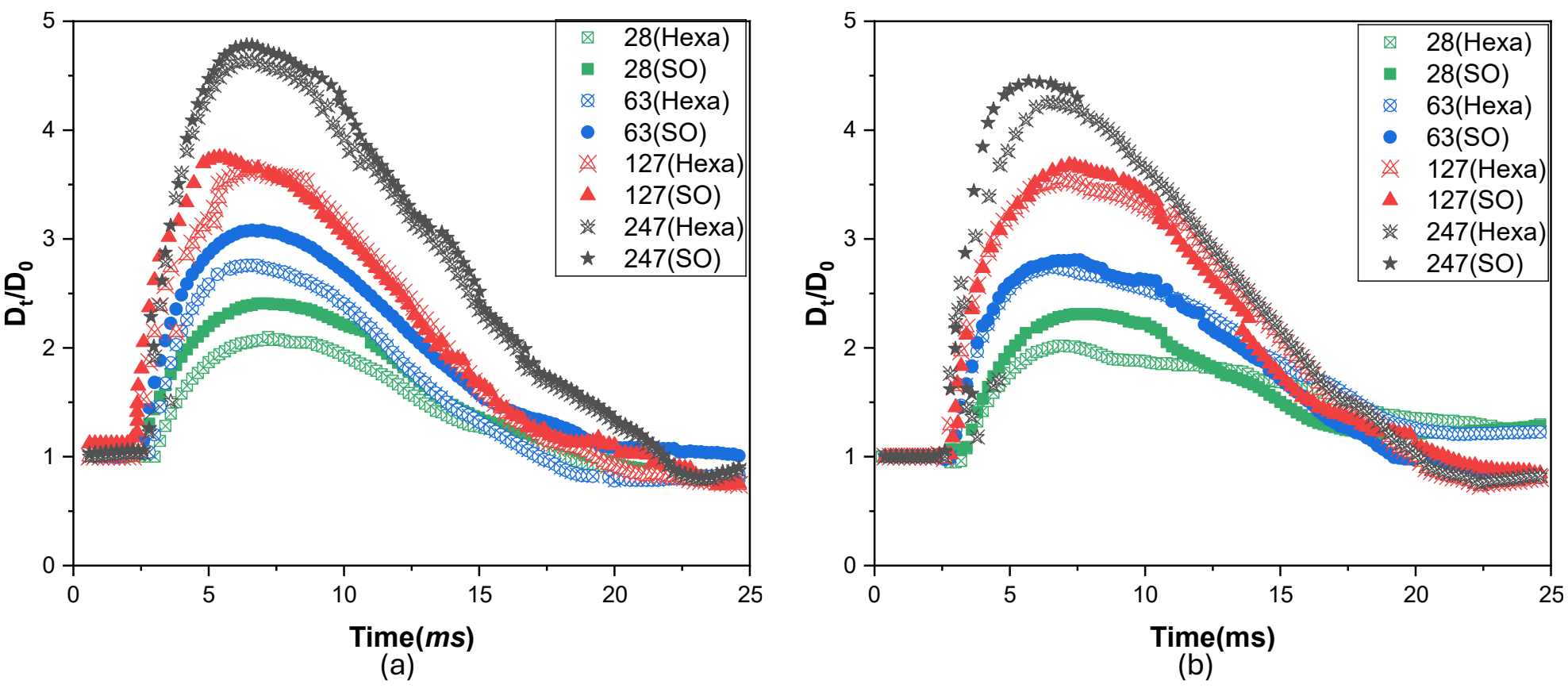

**Figure 13**. Time development of the diameters of the hitting droplet lamellas for the two different surfaces of PDMS-OTS (a) 5µm and (b) 20µm post spacing when coated with SO-5cSt and hexadecane for different Weber numbers.

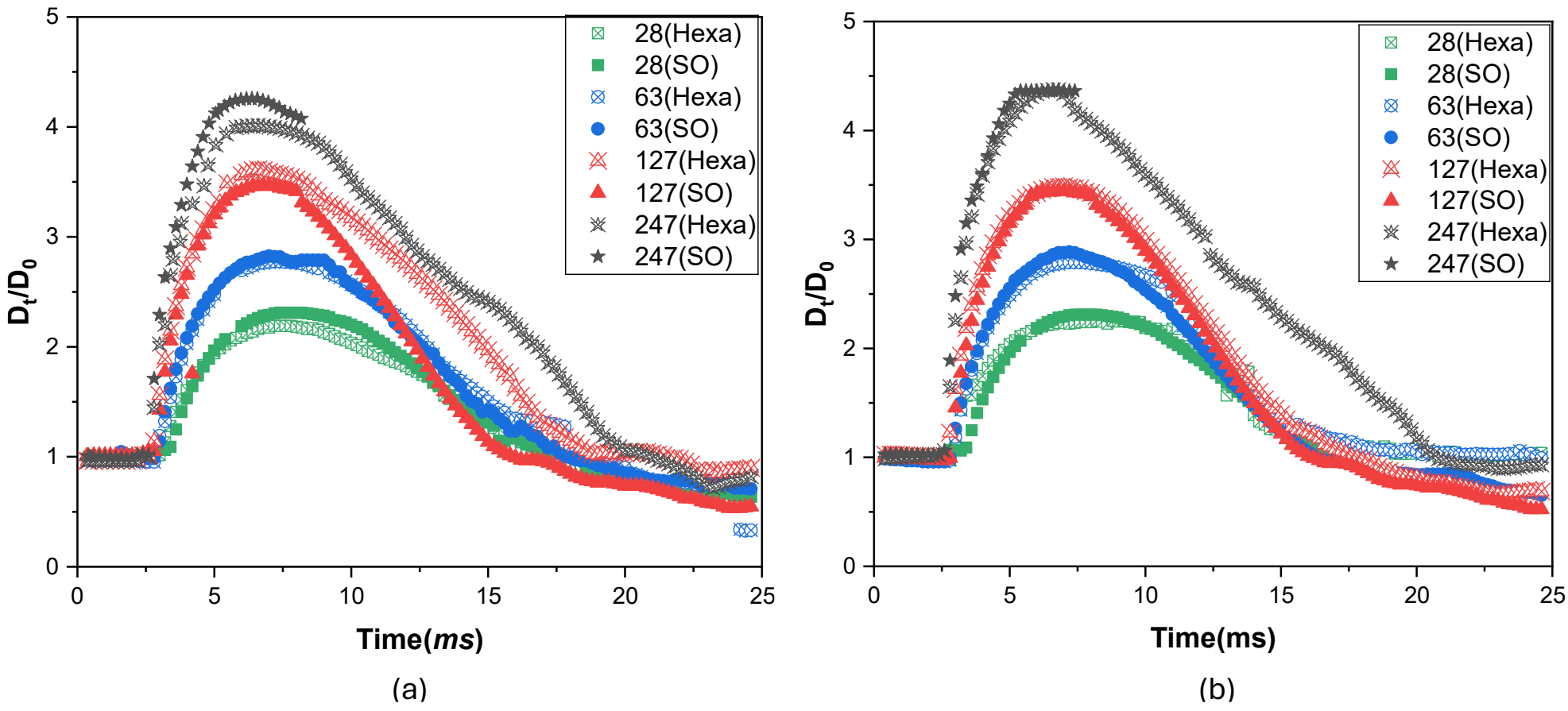

**Figure 14**. Time development of the diameters of the hitting droplet lamellas for the two different surfaces of PDMS-OTS (a) 5µm and (b) 20µm post spacing when absorbed with SO-5cSt and hexadecane for different Weber numbers.

Although this behaviour aligns with established findings in droplet dynamics, its interpretation on lubricant-coated textured surfaces warrants further analysis using a scaling-based approach. The dynamic contact angle at the three-phase composite interface strongly influences the spreading behaviour of droplets on textured PDMS-OTS surfaces coated and absorbed with lubricants. These experimental surfaces possess low solid fractions, making the composite interface prone to disruption under dynamic conditions such as impact. This results in a transition from the Cassie-Baxter to the Wenzel wetting state, increasing the solid-liquid contact area and thereby enhancing viscous dissipation, which significantly affects droplet spreading. To quantify spreading, the non-dimensional maximum spreading factor, this model predicts that the maximum spreading factor $\beta_{max}$ scales with the Weber number as $\beta_{max} \sim We^{\frac{1}{4}}$, which has been reported in experiments[59–61] for the low-viscosity liquids.[6] Kim et al. highlighted that the scaling law for maximum spreading can differ between superhydrophobic surfaces and LIS. Specifically, for LIS, the scaling is expressed, $\beta_{max} = \left[1 + \left(\frac{t}{h}\right)\left(\frac{\mu_w}{\mu_o}\right)\right]^{0.5} We^{\frac{1}{4}}$, where t is the oil film thickness, and h is the thickness of the maximum spreading droplet.. Small viscous-oil correction that only becomes important when the oil layer is a non-negligible fraction of the pancake thickness (i.e.$\left(\frac{t}{h}\right)$ or when $\mu_o$ is much smaller than $\mu_w$). In our case $t \ll h$ and , $\mu_o$ is comparable to or larger than $\mu_w$, so the prefactor ≈1. Thus the for the SLIPs and LIS with less thickness and low viscosity oil fall on the same slope $We^{\frac{1}{4}}$. The maximum spreading factor follows a power-law relationship with the Weber number, expressed as $\beta_{max} \sim We^{\alpha}$ . When the graph was plotted according to the experimental data, On textured PDMS-OTS when coated or absorbed with lubricant surfaces, the maximum spreading of the droplet results in good agreement with the corresponding slope with a value of α ≈ 0.30 (Shown in figure 15) $\beta_{max} \sim We^{\frac{1}{4}}$ [12,62]. The exponent values obtained in our study agree with those reported in previous experimental studies.[12,62]

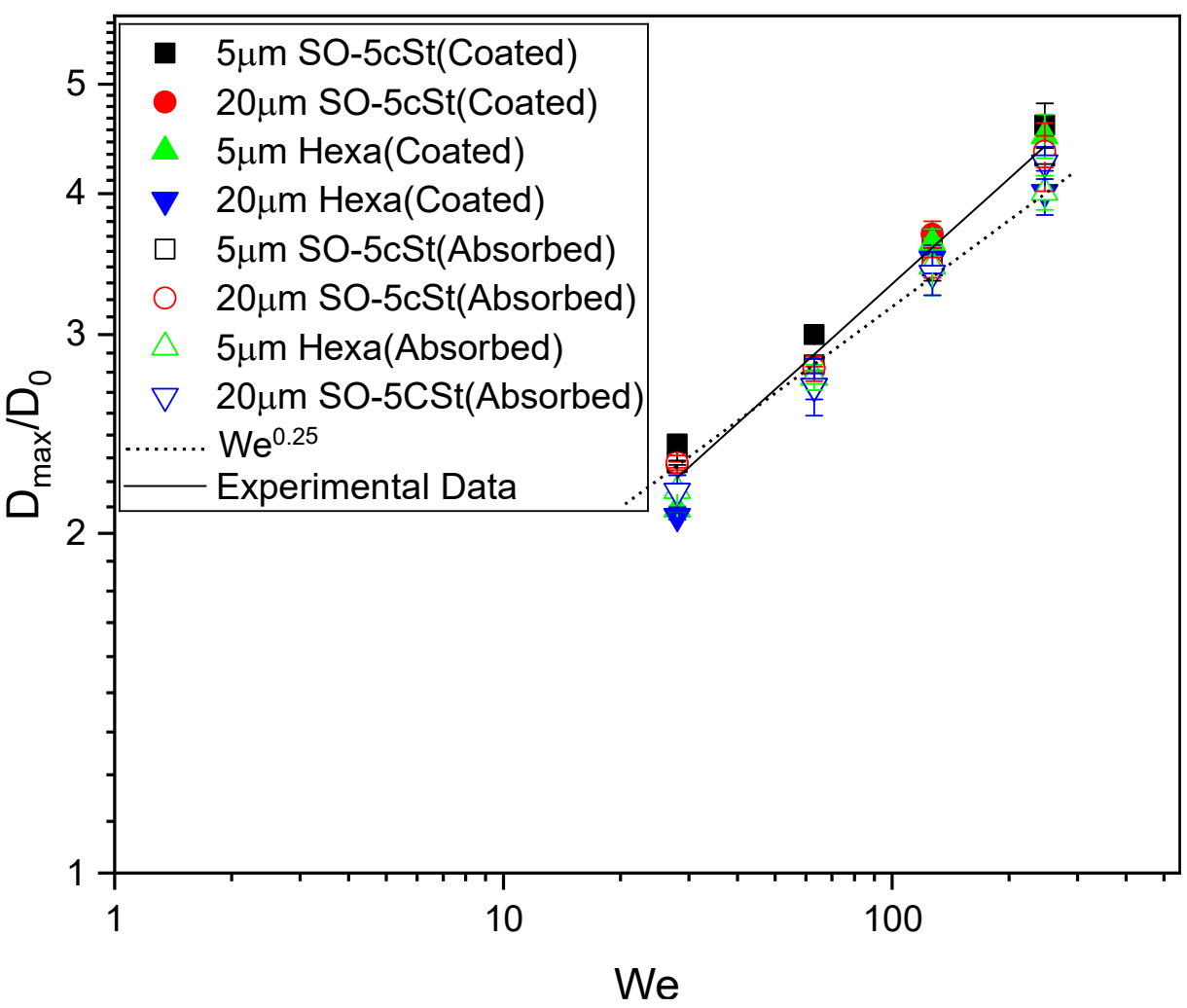

**Figure 15**. Normalized maximum spreading diameter $D_{max}/D$ as a function of the Weber number. Error bars indicate the standard deviation based on four independent measurements.

Figure 16 presents the regime map summarizing the droplet impact outcomes for textured

PDMS-OTS surfaces with 5μm and 20μm post-spacings, for coated and absorbed with SO-5cSt and hexadecane, across the entire range of Weber numbers. The map categorizes the impact behaviours, deposition, partial rebound, and full rebound across a range of Weber numbers and wettability parameter (ratio of $(180^{\circ}/CAH)$). (The detailed value of the weetabilty parameter is shown in the supporting information Table S10). An important observation was made 0)when comparing textured PDMS-OTS samples prepared using two different lubricant application methods. In the coated samples, the lubricant layer was coated on the surface, but the coated lubricant gradually gets absorbed into the uppermost layer of the PDMS-OTS of the PDMS matrix over time due to the porosity. This process left sufficient oil between the micro-posts, facilitating droplet mobility and promoting distinct bouncing regimes after impact. On the other hand, in samples allowed to absorb lubricant overnight, the oil was not only retained between the posts but also absorbed into the bulk of the PDMS-OTS substrate. This dual retention led to a more stable lubricating environment, reducing interfacial friction and enhancing droplet mobility. Notably, even hexadecane-infused samples, which typically exhibit weaker surface affinity, demonstrated reduced frictional resistance when prepared via absorption due to the presence of lubricant in both the bulk and inter-post regions.

When SO-5cSt oil is absorbed into the textured PDMS-OTS surface, it forms a thin, stable film that covers not only the interstitial regions but also the top surfaces of the micro-posts. This creates a continuous, low-friction boundary layer that resembles a Van der Waals-type SLIPs (Slippery Liquid-Infused Porous Surface), where the droplet interacts with a homogeneous lubricant interface, minimizing contact line pinning. In contrast, when a textured PDMS-OTS surface absorbed with hexadecane, it remained primarily within the texture and bulk PDMS matrix but did not rise to cover the tops of the microstructures. As a result, the droplet partly contacts the bare PDMS-OTS surface and oil layer during impact, forming a type of SLIPs that can be classified as non-Van der Waals. Such surfaces provide reduced lubrication, allowing increased solid (PDMS)-liquid (Water) interactions and increasing the likelihood of partial rebound or pinning.

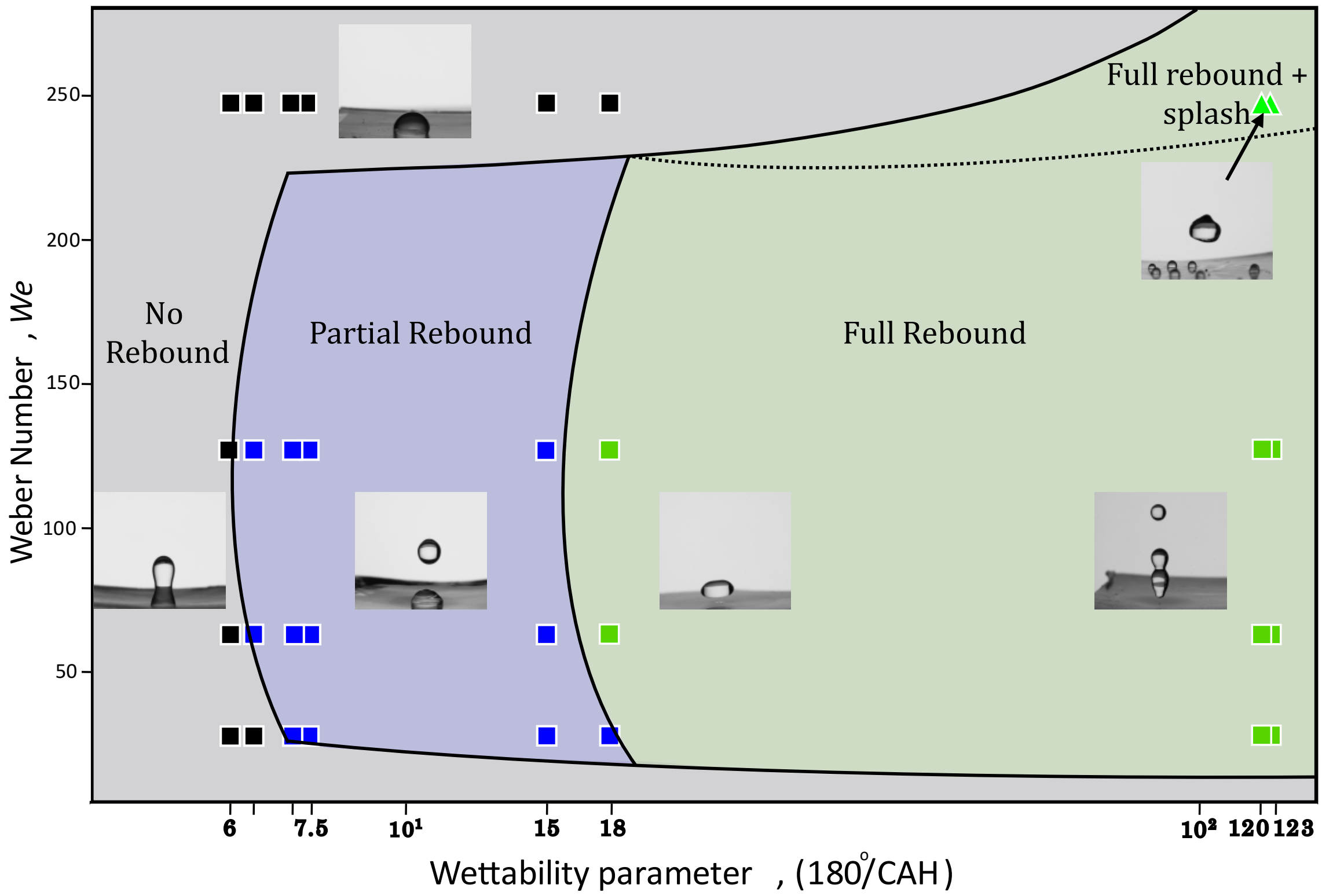

**Figure 16.** Schematic regime map of the outcome of drop impact at the end of the receding cycle for textured PDMS-OTS surface 5 and 20µm when coated and absorbed (separately) with So-5cSt (red square) and hexadecane (black square).

To evaluate the durability of the SLIPs surfaces, repeated droplet-impact experiments were performed on 20µm post-spacing samples at a Weber number of 127 for both SO–5 cSt- and hexadecane-adsorbed surfaces, as both surfaces exhibited complete rebound under these conditions. The impacts were applied at the same location to assess durability and the oil retention capacity of the textured surfaces. For Van der Waals-type SLIPs (SO-5cSt absorbed textured PDMS), complete droplet rebound was maintained for up to 17-20 impact cycles, after which the behavior transitioned to partial rebound. In contrast, for non-Van der Waals-type SLIPs (hexadecane absorbed textured PDMS), this transition occurred much earlier, within 4-8 cycles (Figure S6, Supporting Information). To quantify lubricant loss, gravimetric measurements were conducted after repeated full-rebound impact events on the same sample. The mass of each sample was recorded before and after a prescribed number of impacts at the same location, enabling a direct correlation between the decline in rebound performance and the reduction in lubricant content. The results showed a progressive decrease in sample mass with increasing impact cycles, indicating continuous lubricant depletion from both surface types. The gravimetric measurements reveal a clear distinction in lubricant depletion behavior between the two systems. Across all trials, the total lubricant loss remains minimal, at the milligram level. In comparison, Van der Waals-SLIPs exhibit a significantly slower reduction in mass, indicating a lower rate of lubricant depletion. This behavior is attributed to cloaking-driven transfer, in which only a thin silicone oil film is deposited on the impacting droplets, resulting in negligible lubricant removal per impact. Conversely, non-Van der Waals SLIPs exhibit a clearly higher rate of mass loss. This is primarily due to displacement and partial mixing of hexadecane with the impacting droplet, as well as during spreading and recoil. Furthermore, shedding during rebound contributes to additional lubricant loss. Table S11 and Figure S7 in the supporting information summarize the percentage change in mass following successive droplet impacts for both Van der Waals-SLIPs and non-Van der Waals-SLIPs. Overall, the experiment highlights the superior interfacial stability and impact-resistance of Van der Waals-SLIPs relative to non-Van der Waals-SLIPs.

## 4. Conclusions

In this study, the effect of surface texture on droplet impact dynamics is investigated by fabricating PDMS surfaces with square micropost arrays with post spacings of 5μm and 20μm using soft lithography. These textured surfaces were then functionalized with OTS to enhance their non-wettability. Following functionalization, the samples were either coated with or allowed to absorb two different lubricants, SO-5cSt and hexadecane. Wettability measurements were conducted on all samples, and droplet impact experiments were performed at various Weber numbers (28, 63, 127, and 247) to assess the influence of texture, surface chemistry, and lubricant interaction on static and dynamic wettability behavior. The droplet impact behaviour on textured PDMS surfaces showed only minor differences between the unfunctionalized samples (Textured PDMS) and the OTS-functionalized samples (Textured PDMS-OTS) in terms of bouncing behaviour. However, significant changes in droplet bouncing behaviour were observed when the OTS-functionalized PDMS textured surfaces were coated with lubricants. The type of lubricant and the spacing between the posts played a crucial role in determining whether the droplet rebounded, partially rebounded, or adhered to the surface.

When the textured PDMS-OTS surfaces were absorbed with lubricants, apparent differences in rebound behaviour were observed. For example, surfaces absorbed with SO-5cSt showed complete rebound across all Weber numbers for both the post spacings, indicating stable lubricant retention and a continuous oil layer at the surface. In contrast, hexadecane-absorbed surfaces showed variable behaviour depending on both the post spacing and the Weber number, with a partial rebound at a lower Weber number and no rebound at higher Weber numbers for the same 5 μm post spacing. This phenomenon was explained based on the balance between impact and capillary pressures. Durability was also assessed by repeating droplet impacts. It was observed that surfaces with SO-5cSt or hexadecane absorbed into the textured PDMS maintained rebound for more cycles than solid textured silicon wafers. This demonstrates that PDMS's ability to retain oil internally helps preserve its water droplet-repelling properties during repeated impacts. This study highlights how surface texture, chemical treatment, and internal oil absorption can significantly influence droplet rebound behaviour. These insights are helpful in developing surfaces for applications such as biofouling, anti-microbial coatings, and anti-icing.

## Author contributions

Shubham Ganar: Designed and performed the experiments; theoretical calculations; data analysis; wrote the original manuscript draft. Deepak J: Performed experiments; reviewed the manuscript. Arindam Das: Conceived and developed the experimental design and overall research concept; reviewed and edited the manuscript; provided supervision throughout the work.

**Conflicts of interest**

There are no conflicts to declare.

**Data availability**

The data supporting this article are included in the supporting information (SI). The supporting information contains additional images.

**Supporting Information**

See the supporting information for the details of the experiment and results. It is divided into sections: 1. Fabrication on textured PDMS surface, 2. Results of wettability measurements, 3. Stability and thermodynamic framework, 4. FESEM Images, 5. Experimental critical *We* for the transition in bouncing phenomena, 6. Regime map, repeatability, and oil retention.

**Acknowledgements**

The authors would like to thank the School of Mechanical Sciences and the Centre of Excellence in Particulates, Colloids, and Interfaces at the Indian Institute of Technology Goa for providing the experimental facility and necessary support to conduct the above work. We would like to thank R. Aakash, an intern at IIT Goa from the National Institute of Technology, Tiruchirapalli. The Science and Engineering Research Board funded this research work with Sanction Order Nos. CRG/2023/008620 and SRG/2019/002011.

## References

(1) A. M. Worthington. On the Forms Assumed by Drops of Liquids Falling Vertically on a Horizontal Plate Author ( s ): A . M . Worthington Source : Proceedings of the Royal Society of London , Vol . 25 ( 1876 - 1877 ), Pp . 261-272 Published by : The Royal Society Stable URL : *Proc. R. Soc. London* **1876**, *25*, 261–272.

(2) Muschi, M.; Brudieu, B.; Teisseire, J.; Sauret, A. Drop Impact Dynamics on Slippery Liquid-Infused Porous Surfaces: Influence of Oil Thickness. *Soft Matter* **2018**, *14* (7), 1100–1107. https://doi.org/10.1039/c7sm02026k.

(3) Lee, E.; Chilukoti, H. K.; Müller-Plathe, F. Suppressing the Rebound of Impacting Droplets from Solvophobic Surfaces by Polymer Additives: Polymer Adsorption and Molecular Mechanisms. *Soft Matter* **2021**, *17* (29), 6952–6963. https://doi.org/10.1039/d1sm00558h.

(4) Jayaprakash, V.; Rufer, S.; Panat, S.; Varanasi, K. K. Enhancing Spray Retention Using Cloaked Droplets to Reduce Pesticide Pollution. *Soft Matter* **2025**, *21* (19). https://doi.org/10.1039/d4sm01496k.

(5) Wang, H.; Lu, H.; Zhao, W. A Review of Droplet Bouncing Behaviors on Superhydrophobic Surfaces: Theory, Methods, and Applications. *Phys. Fluids* **2023**, *35* (2), 021301 (1-23). https://doi.org/10.1063/5.0136692.

(6) Josserand, C.; Thoroddsen, S. T. Drop Impact on a Solid Surface. *Annu. Rev. Fluid Mech.* **2016**, *48* (September 2015), 365–391. https://doi.org/10.1146/annurev-fluid-122414-034401.

(7) Yarin, A. L. Drop Impact Dynamics: Splashing, Spreading, Receding, Bouncing.. *Annu. Rev. Fluid Mech.* **2006**, *38*, 159–192. https://doi.org/10.1146/annurev.fluid.38.050304.092144.

(8) Quéré, D. Leidenfrost Dynamics. *Annu. Rev. Fluid Mech.* **2013**, *45*, 197–215. https://doi.org/10.1146/annurev-fluid-011212-140709.

(9) Tran, T.; Staat, H. J. J.; Susarrey-Arce, A.; Foertsch, T. C.; van Houselt, A.; Gardeniers, H. J. G. E.; Prosperetti, A.; Lohse, D.; Sun, C. Droplet Impact on Superheated Micro-Structured Surfaces. *Soft Matter* **2013**, *9* (12), 3272–3282. https://doi.org/10.1039/C3SM27643K.

(10) Zhang, H.; Ahumada Lazo, J.; Sarwar, M. S. Bin; Liu, Y. Revealing the Dynamic and Thermal Behaviors of Supercooled Droplet Impinging on Surfaces with Varying Wettability. *Soft Matter* **2025**, *21* (45), 8655–8668. https://doi.org/10.1039/d5sm00761e.

(11) Sharma, S.; Pinto, R.; Saha, A.; Chaudhuri, S.; Basu, S. On Secondary Atomization and Blockage of Surrogate Cough Droplets in Single- And Multilayer Face Masks. *Sci. Adv.* **2021**, *7* (10), 1–12. https://doi.org/10.1126/sciadv.abf0452.

(12) Khojasteh, D.; Kazerooni, M.; Salarian, S.; Kamali, R. Droplet Impact on Superhydrophobic Surfaces: A Review of Recent Developments. *J. Ind. Eng. Chem.* **2016**, *42*, 1–14. https://doi.org/10.1016/j.jiec.2016.07.027.

(13) Yu, X.; Zhang, Y.; Hu, R.; Luo, X. Water Droplet Bouncing Dynamics. *Nano Energy* **2021**, *81* (November 2020), 105647. https://doi.org/10.1016/j.nanoen.2020.105647.

(14) Patankar, N. A. On the Modeling of Hydrophobic Contact Angles on Rough Surfaces. *Langmuir* **2003**, *19* (4), 1249–1253. https://doi.org/10.1021/la026612+.

(15) Patankar, N. A. Transition between Superhydrophobic States on Rough Surfaces. *Langmuir* **2004**, *20* (17), 7097–7102. https://doi.org/10.1021/la049329e.

(16) Nonomura, Y.; Tanaka, T.; Mayama, H. Penetration Behavior of a Water Droplet into a Cylindrical Hydrophobic Pore. *Langmuir* **2016**, *32* (25), 6328–6334. https://doi.org/10.1021/acs.langmuir.6b01509.

(17) Jung, Y. C.; Bhushan, B. Dynamic Effects of Bouncing Water Droplets on Superhydrophobic Surfaces. *Langmuir* **2008**, *24* (12), 6262–6269. https://doi.org/10.1021/la8003504.

(18) Mitra, S.; Vo, Q.; Tran, T. Bouncing-to-Wetting Transition of Water Droplets Impacting Soft Solids. *Soft Matter* **2021**, *17* (24), 5969–5977. https://doi.org/10.1039/d1sm00339a.

(19) Wang, L. Z.; Zhou, A.; Zhou, J. Z.; Chen, L.; Yu, Y. S. Droplet Impact on Pillar-Arrayed Non-Wetting Surfaces. *Soft Matter* **2021**, *17* (24), 5932–5940. https://doi.org/10.1039/d1sm00354b.

(20) Wu, J.; Zhang, L.; Lu, Y.; Yu, Y. Droplet Impinging on Sparse Micropillar-Arrayed Non-Wetting Surfaces. *Phys. Fluids* **2024**, *36* (9). https://doi.org/10.1063/5.0226032.

(21) Zhang, L.; Wu, J.; Lu, Y.; Yu, Y. Droplets Impact on Sparse Microgrooved Non-Wetting Surfaces. *Sci. Rep.* **2025**, *15* (1), 2918. https://doi.org/10.1038/s41598-025-87294-z.

(22) Villegas, M.; Zhang, Y.; Abu Jarad, N.; Soleymani, L.; Didar, T. F. Liquid-Infused Surfaces: A Review of Theory, Design, and Applications. *ACS Nano* **2019**, *13* (8), 8517–8536. https://doi.org/10.1021/acsnano.9b04129.

(23) Li, J.; Zhou, Z.; Jiao, X.; Guo, Z.; Fu, F. Bioinspired Lubricant-Infused Porous Surfaces: A Review on Principle, Fabrication, and Applications. *Surfaces and Interfaces* **2024**, *53* (July). https://doi.org/10.1016/j.surfin.2024.105037.

(24) Zhang, D.; Xia, Y.; Chen, X.; Shi, S.; Lei, L. PDMS-Infused Poly(High Internal Phase Emulsion) Templates for the Construction of Slippery Liquid-Infused Porous Surfaces with Self-Cleaning and Self-Repairing Properties. *Langmuir* **2019**, *35* (25), 8276–8284. https://doi.org/10.1021/acs.langmuir.9b01115.

(25) Li, J.; Lu, B.; Cheng, Z.; Cao, H.; An, X. Designs and Recent Progress of "Pitcher Plant Effect" Inspired Ultra-Slippery Surfaces: A Review. *Prog. Org. Coatings* **2024**, *191* (April), 108460. https://doi.org/10.1016/j.porgcoat.2024.108460.

(26) Ganar, S. S.; Das, A. Experimental Insights into Droplet Behavior on Van Der Waals and Non-Van Der Waals Liquid-Impregnated Surfaces. **2024**, No. December 2022. https://doi.org/10.1063/5.0236861.

(27) Yeganehdoust, F.; Attarzadeh, R.; Dolatabadi, A.; Karimfazli, I. A Comparison of Bioinspired Slippery and Superhydrophobic Surfaces: Micro-Droplet Impact. *Phys. Fluids* **2021**, *33* (2).

https://doi.org/10.1063/5.0035556.

(28) Wong, T.; Kang, S. H.; Tang, S. K. Y.; Smythe, E. J.; Hatton, B. D.; Grinthal, A.; Aizenberg, J. Bioinspired Self-Repairing Slippery Surfaces with Pressure-Stable Omniphobicity. *Nature* **2011**, *477* (7365), 443–447. https://doi.org/10.1038/nature10447.

(29) Rapoport, L.; Solomon, B. R.; Varanasi, K. K. Mobility of Yield Stress Fluids on Lubricant-Impregnated Surfaces. *ACS Appl. Mater. Interfaces* **2019**, *11* (17), 16123–16129. https://doi.org/10.1021/acsami.8b21478.

(30) Nayse, A. K.; Mund, A.; Ganar, S.; Das, A. Design and Fabrication of Industrially Scalable Low-Cost Liquid Impregnated Omni-Phobic Surfaces with Extreme Hydratephobic Properties. *Surfaces and Interfaces* **2025**, *70* (May). https://doi.org/10.1016/j.surfin.2025.106744.

(31) Lee, C.; Kim, H.; Nam, Y. Drop Impact Dynamics on Oil-Infused Nanostructured Surfaces. *Langmuir* **2014**, *30* (28), 8400–8407. https://doi.org/10.1021/la501341x.

(32) Chen, F.; Wang, Y.; Tian, Y.; Zhang, D.; Song, J.; Crick, C. R.; Carmalt, C. J.; Parkin, I. P.; Lu, Y. Robust and Durable Liquid-Repellent Surfaces. *Chem. Soc. Rev.* **2022**, *51* (20), 8476–8583. https://doi.org/10.1039/d0cs01033b.

(33) Ganar, S. S.; J, D.; Das, A. High-Speed Imagery Analysis of Droplet Impact on van Der Waals and Non-van Der Waals Soft Oil-Infused Surfaces. *Langmuir* **2025**. https://doi.org/10.1021/acs.langmuir.5c02765.

(34) Das, A.; Farnham, T. A.; Bengaluru Subramanyam, S.; Varanasi, K. K. Designing Ultra-Low Hydrate Adhesion Surfaces by Interfacial Spreading of Water-Immiscible Barrier Films. *ACS Appl. Mater. Interfaces* **2017**, *9* (25), 21496–21502. https://doi.org/10.1021/acsami.7b00223.

(35) Ghosh, N.; Bajoria, A.; Vaidya, A. A. Surface Chemical Modification of Poly(Dimethylsiloxane)-Based Biomimetic Materials: Oil-Repellent Surfaces. *ACS Appl. Mater. Interfaces* **2009**, *1* (11), 2636–2644. https://doi.org/10.1021/am9004732.

(36) Cao, Y.; Jana, S.; Tan, X.; Bowen, L.; Zhu, Y.; Dawson, J.; Han, R.; Exton, J.; Liu, H.; McHale, G.; Jakubovics, N. S.; Chen, J. Antiwetting and Antifouling Performances of Different Lubricant-Infused Slippery Surfaces. *Langmuir* **2020**, *36* (45), 13396–13407. https://doi.org/10.1021/acs.langmuir.0c00411.

(37) Qin, D.; Xia, Y.; Whitesides, G. M. Soft Lithography for Micro- and Nanoscale Patterning. *Nat. Protoc.* **2010**, *5* (3), 491–502. https://doi.org/10.1038/nprot.2009.234.

(38) Vaillard, A. S.; Saget, M.; Braud, F.; Lippert, M.; Keirsbulck, L.; Jimenez, M.; Coffinier, Y.; Thomy, V. Highly Stable Fluorine-Free Slippery Liquid Infused Surfaces. *Surfaces and Interfaces* **2023**, *42* (PA), 103296. https://doi.org/10.1016/j.surfin.2023.103296.

(39) Dawson, J.; Coaster, S.; Han, R.; Gausden, J.; Liu, H.; McHale, G.; Chen, J. Dynamics of Droplets Impacting on Aerogel, Liquid Infused, and Liquid-Like Solid Surfaces. *ACS Appl. Mater. Interfaces* **2023**, *15* (1), 2301–2312. https://doi.org/10.1021/acsami.2c14483.

(40) Alizadeh, A.; Bahadur, V.; Shang, W.; Zhu, Y.; Buckley, D.; Dhinojwala, A.; Sohal, M. Influence of Substrate Elasticity on Droplet Impact Dynamics. *Langmuir* **2013**, *29* (14), 4520–4524. https://doi.org/10.1021/la304767t.

(41) Chen, L.; Bonaccurso, E.; Deng, P.; Zhang, H. Droplet Impact on Soft Viscoelastic Surfaces. *Phys. Rev. E* **2016**, *94* (6), 1–9. https://doi.org/10.1103/PhysRevE.94.063117.

(42) Kanungo, M.; Mettu, S.; Law, K. Y.; Daniel, S. Effect of Roughness Geometry on Wetting and Dewetting of Rough PDMS Surfaces. *Langmuir* **2014**, *30* (25), 7358–7368. https://doi.org/10.1021/la404343n.

(43) Smith, J. D.; Meuler, A. J.; Bralower, H. L.; Venkatesan, R.; Subramanian, S.; Cohen, R. E.; McKinley, G. H.; Varanasi, K. K. Hydrate-Phobic Surfaces: Fundamental Studies in Clathrate Hydrate Adhesion Reduction. *Phys. Chem. Chem. Phys.* **2012**, *14* (17), 6013–6020. https://doi.org/10.1039/c2cp40581d.

(44) Smith, J. D.; Dhiman, R.; Anand, S.; Reza-Garduno, E.; Cohen, R. E.; McKinley, G. H.; Varanasi, K. K. Droplet Mobility on Lubricant-Impregnated Surfaces. *Soft Matter* **2013**, *9* (6), 1772–1780. https://doi.org/10.1039/c2sm27032c.

(45) Israelachvili, J. N. Van Der Waals Forces between Particles and Surfaces. In *Intermolecular and Surface Forces*; Elsevier, 2011; pp 253–289. https://doi.org/10.1016/B978-0-12-391927-4.10013-1.

(46) Polytechnique, E.; Cedex, P.; Seiwert, J.; Clanet, C.; Quéré, D. Coating of a Textured Solid. *J. Fluid Mech.* **2011**, *669*, 55–63. https://doi.org/10.1017/S0022112010005951.

(47) Li, J.; Kleintschek, T.; Rieder, A.; Cheng, Y.; Baumbach, T.; Obst, U.; Schwartz, T.; Levkin, P. A. Hydrophobic Liquid-Infused Porous Polymer Surfaces for Antibacterial Applications. *ACS Appl. Mater. Interfaces* **2013**, *5* (14), 6704–6711. https://doi.org/10.1021/am401532z.

(48) Van de Velde, P.; Fabre-Parras, N.; Josserand, C.; Duprat, C.; Protière, S. Spreading and Absorption of a Drop on a Swelling Surface. *Europhys. Lett.* **2023**, *144* (3), 33001. https://doi.org/10.1209/0295-5075/ad0eed.

(49) Ganar, S. S.; Das, A. Unraveling the Interplay of Leaf Structure and Wettability: A Comparative Study on Superhydrophobic Leaves of Cassia Tora, Adiantum Capillus-Veneris, and Bauhinia Variegata. *Phys. Fluids* **2023**, *35* (11), 1–38. https://doi.org/10.1063/5.0172707.

(50) Yada, S.; Lacis, U.; Van Der Wijngaart, W.; Lundell, F.; Amberg, G.; Bagheri, S. Droplet Impact on Asymmetric Hydrophobic Microstructures. *Langmuir* **2022**, *38* (26), 7956–7964. https://doi.org/10.1021/acs.langmuir.2c00561.

(51) Guo, C.; Zhao, D.; Sun, Y.; Wang, M.; Liu, Y. Droplet Impact on Anisotropic Superhydrophobic Surfaces. *Langmuir* **2018**, *34* (11), 3533–3540. https://doi.org/10.1021/acs.langmuir.7b03752.

(52) Xu, L. Liquid Drop Splashing on Smooth, Rough, and Textured Surfaces. *Phys. Rev. E* **2007**, *75* (5), 056316. https://doi.org/10.1103/PhysRevE.75.056316.

(53) Marmanis, H.; Thoroddsen, S. T. Scaling of the Fingering Pattern of an Impacting Drop. *Phys. Fluids* **1996**, *8* (6), 1344–1346. https://doi.org/10.1063/1.868941.

(54) Gidreta, B. T.; Huang, M.; Daniel, D.; Adera, S. Drop Impact Dynamics on Hierarchically Textured Lubricant-Infused Surfaces. *Phys. Rev. Fluids* **2025**, *10* (1), 013604. https://doi.org/10.1103/PhysRevFluids.10.013604.

(55) Allen, R. F. The Role of Surface Tension in Splashing. *J. Colloid Interface Sci.* **1975**, *51* (2), 350–351. https://doi.org/10.1016/0021-9797(75)90126-5.

(56) Huang, X.; Wan, K. T.; Taslim, M. E. Axisymmetric Rim Instability of Water Droplet Impact on a Super-Hydrophobic Surface. *Phys. Fluids* **2018**, *30* (9). https://doi.org/10.1063/1.5039558.

(57) Liu, D.; Yin, H.; Wu, Z.; Luo, X. Retraction Dynamics of an Impacting Droplet on a Rotating Surface. *Phys. Rev. Fluids* **2025**, *10* (3), 033602. https://doi.org/10.1103/PhysRevFluids.10.033602.

(58) Wang, F.; Fang, T. Retraction Dynamics of Water Droplets after Impacting upon Solid Surfaces from Hydrophilic to Superhydrophobic. *Phys. Rev. Fluids* **2020**, *5* (3), 033604. https://doi.org/10.1103/PhysRevFluids.5.033604.

(59) Clanet, C.; Béguin, C.; Richard, D.; Quéré, D. Maximal Deformation of an Impacting Drop. *J. Fluid Mech.* **2004**, *517*, 199–208. https://doi.org/10.1017/S0022112004000904.

(60) Bennett, T.; Poulikakos, D. Splat-Quench Solidification: Estimating the Maximum Spreading of a Droplet Impacting a Solid Surface. *J. Mater. Sci.* **1993**, *28* (4), 963–970. https://doi.org/10.1007/BF00400880.

(61) Bartolo, D.; Josserand, C.; Bonn, D. Retraction Dynamics of Aqueous Drops upon Impact on Non-Wetting Surfaces. *J. Fluid Mech.* **2005**, *545*, 329–338. https://doi.org/10.1017/S0022112005007184.

(62) Ding, S.; Hu, Z.; Dai, L.; Zhang, X.; Wu, X. Droplet Impact Dynamics on Single-Pillar Superhydrophobic Surfaces. *Phys. Fluids* **2021**, *33* (10), 102108. https://doi.org/10.1063/5.0066366.

*Supporting Information*

# High-Speed Imagery Analysis of Droplet Impact on Van der Waals and Non-Van der Waals Soft-Textured Oil-Infused Surfaces.

Shubham S. Ganar[1], Deepak J.[1] and Arindam Das[1]*
[1]School of Mechanical Sciences, Indian Institute of Technology (IIT) Goa, GEC Campus, Farmagudi, Ponda, Goa 403401, India

**Corresponding Author:** Arindam Das*, Associate Professor, School of Mechanical Sciences, Indian Institute of Technology (IIT) Goa, Email: arindam@iitgoa.ac.in,

## 1. Fabrication on textured PDMS surface.

To prepare textured PDMS samples for droplet impact experiments, we first fabricated microtextured silicon surfaces with 10 μm square posts, interpost spacings of 5 and 20 μm, and a height of 10 μm using standard lithography. Silicon wafers were coated with Shipley S1818 photoresist and exposed to 405 nm UV light through a chrome mask (Advanced Reproductions Corporation). The photoresist was developed in a 1:1 mixture of DI water and Microdev solution (Dow Chemicals). Etching was performed to a depth of 10 μm using an inductively coupled plasma reactor (Surface Technology Systems). Surface profiles were measured with an optical profiler (CCI HD, Taylor Hobson). Residual photoresist was removed using a piranha solution (3:1 sulfuric acid to hydrogen peroxide).

Polydimethylsiloxane (PDMS) is a flexible polymer widely used for fabricating microstructures due to its ease of moulding. In this study, soft surfaces were prepared using PDMS (Sylgard 184, Dow Corning, Wiesbaden, Germany). Liquid PDMS was mixed with a curing agent in a 10:1 ratio (PDMS to curing agent) using a mechanical stirrer for 5 minutes.[1] This process introduced air bubbles, which were removed by placing the mixture under vacuum for 30 minutes.[2] The degassed PDMS was then used to prepare textured PDMS surfaces through a soft lithography process.[3] For textured surfaces, the degassed PDMS was poured onto silicon wafers coated with a fluorosilane layer. The wafers were placed inside a custom mold, and the PDMS was cured at room temperature for 24 hours to form a negative mold of the textured surface. The cured PDMS mold was then coated with a fluorosilane layer to prevent sticking. A fresh batch of degassed PDMS was poured onto this mold and cured under the same conditions. Once removed, the PDMS formed a positive mold that reproduced the original textured silicon surface. This method enabled clean replication and easy demolding of PDMS layers, which were later used in droplet impact studies. A schematic of the preparation process is shown in Figure S1.

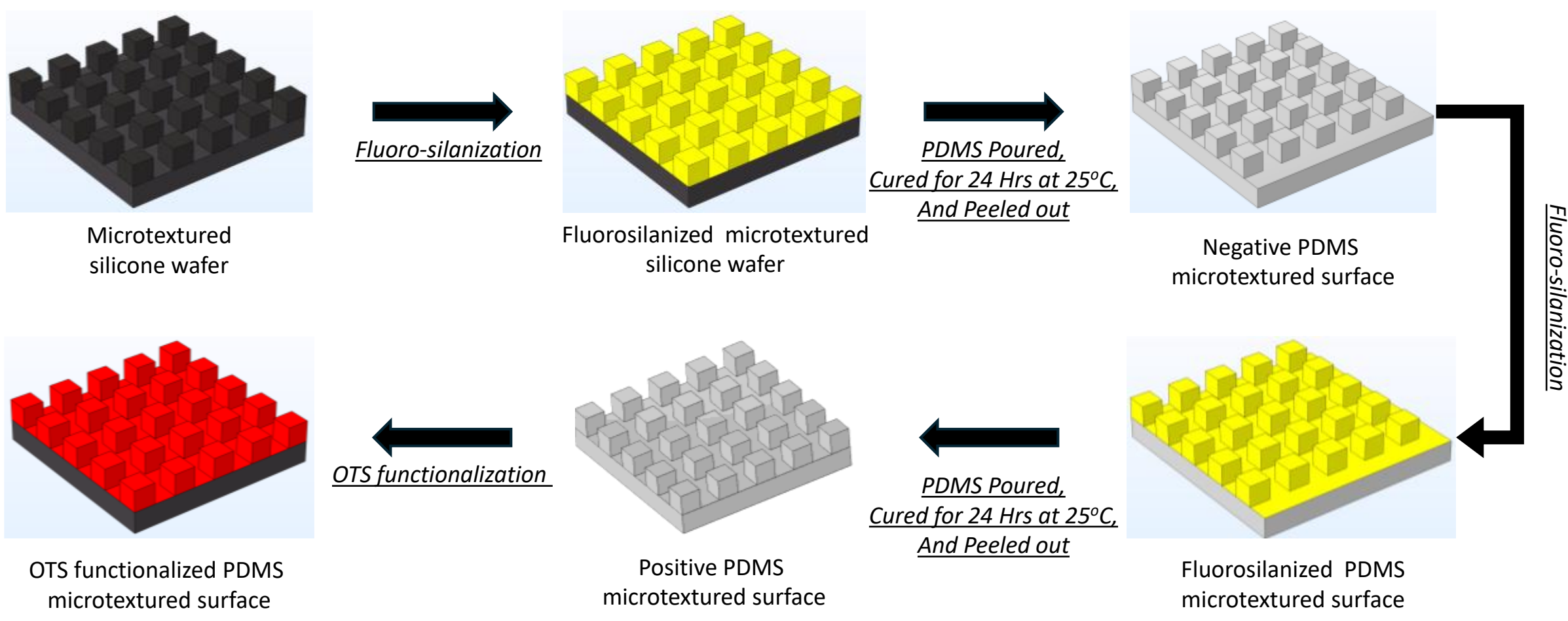

**Figure S1.** The schematic diagram outlines the procedures for preparing textured PDMS for droplet impact tests.

### 1.1. OTS functionalization.

Before functionalization, the samples were plasma-cleaned for 2 minutes to activate the surfaces. A reactive solution was prepared by mixing 75 mL of toluene with 250 μL of octadecyltrichlorosilane (OTS). Separately, a water-in-toluene emulsion was made by combining 325 μL of deionized water (resistivity 18.2 MΩ, Millipore) with 50 mL of toluene, followed by high-energy probe sonication (750 W, Sonics) for 90 seconds. This emulsion was then added to the OTS–toluene solution and further mixed using bath sonication (Branson) for 2 minutes to ensure uniformity. The samples were immersed in the silanization solution for at least 20 minutes. After functionalization, they were rinsed thoroughly with acetone and isopropanol to remove unreacted OTS and byproducts.[4]

**Table S1.** Physical Properties of Lubricant.

| | **SO-5*cSt*** | **Hexadecane** |
|---|---|---|
| Kinematic viscosity (*cSt*) | 5 | 4.3 |
| Specific gravity | 0.91 | 0.71 |
| Dynamic viscosity(*mPa-s*) | 4.57 | 3.06 |
| Surface tension (*mN/m*) | 19.7 | 27.47 |

## 2. Results of wettability measurements

After sample preparation, wettability measurements were performed. All samples were mounted on a goniometer (Rame Hart, Model 500) to measure equilibrium, advancing, and receding contact angles, as well as droplet roll-off angles. A monochrome video camera attached to the goniometer was used to capture droplet images. Deionized (DI) water droplets of 8 μL volume were placed vertically on the test surfaces for contact angle measurements. Ten measurements were taken for each sample type, covering five different locations per sample.[5] Experiments were carried out at 24 °C and 75% relative humidity. Contact angle hysteresis (CAH) was determined using the drop volume-change method.[5] A needle was positioned near the surface to add water gradually until the advancing contact angle was reached, just before the three-phase contact line (TPCL) moved forward. The receding contact angle was recorded during suction when the TPCL began to retract. CAH was calculated as the difference between the advancing and receding angles. Droplet roll-off angles were measured by placing a sessile droplet on the surface and tilting the goniometer stage until the droplet rolled off. The tilt angle at this point was recorded as the roll-off angle.

**Table S2.** Wettability measurements for water of Textured PDMS-OTS 5 and 20 μm samples coated with SO-5cSt and hexadecane lubricant, respectively. All measurements are in Degrees(°).

| | 5μm | | 20μm | |
|---|---|---|---|---|
| | Hexadecane | SO-5cSt | Hexadecane | SO-5cSt |
| **Functionalization** | PDMS-OTS | PDMS-OTS | PDMS-OTS | PDMS-OTS |
| **Eq. CA** | $122 \pm 1$ | $120 \pm 1$ | $119 \pm 3$ | $113 \pm 3$ |
| **Advancing CA** | $124 \pm 2$ | $142 \pm 2$ | $121 \pm 3$ | $114 \pm 3$ |
| **Receding CA** | $96 \pm 2$ | $118 \pm 2$ | $91 \pm 4$ | $89 \pm 3$ |
| **CAH** | $28 \pm 4$ | $24 \pm 4$ | $30 \pm 7$ | $25 \pm 6$ |

**Table S3.** Wettability measurements for water of Textured PDMS-OTS 5 and 20 μm samples absorbed with SO-5cSt and hexadecane lubricant, respectively. All measurements are in Degrees(°).

| | 5μm | | 20μm | |
|---|---|---|---|---|
| | Hexadecane | SO-5cSt | Hexadecane | SO-5cSt |
| **Functionalization** | PDMS-OTS | PDMS-OTS | PDMS-OTS | PDMS-OTS |
| **Eq. CA** | 89±1 | 90 ± 1 | 85±7 | 92 ± 1 |
| **Advancing CA** | 92±2 | 91 ± 0.5 | 92±1 | 93±0.5 |
| **Receding CA** | 82±2 | 89.5±0.5 | 80±1 | 91.5 ± 0.5 |
| **CAH** | 10±4 | 1.5 ± 1 | 12 ± 2 | 1.4 ± 1 |

**Table S4.** Wettability of measurements of water, SO-5cSt, and hexadecane on smooth PDMS-OTS samples, in air and water environments, respectively. All measurements are in Degrees(°).

| **Liquid** | $Eq.CA_{(a)}$ | $\theta_{adv,OS(a)}$ | $\theta_{rec,OS(a)}$ | $Eq.CA_{(w)}$ | $\theta_{adv,OS(w)}$ | $\theta_{rec,OS(w)}$ |
|---|---|---|---|---|---|---|
| **Water** | 113±4 | 112±1.5 | 97±2.5 | NA | NA | NA |
| **Silicone Oil(5*cSt*)** | 1 ± 0.5 | 4 ±2 | | 30 ± 2 | 36 ±2 | 5 ± 4 |
| **Hexadecane** | 41 ± 4 | 45 ± 4 | | 33 ± 6 | 36 ± 3 | 25 ± 3 |

## 3. Stability and thermodynamic framework.

Figure S2. Shows the 3D structure of the square post-spacing sample. The Figure S2(1). The dimensions of the post spacing are given, where a = the size of the post, b = the distance between two consecutive posts (post spacing), and h = the height of the post. The mathematical representation as a*b*h. Our experiment uses two different post spacings: i.e., b = 5, and 20$\mu m$. Figure S2 (2) shows solid fraction φ (the ratio of emerged surface area to projected surface area). The solid fraction can be calculated by $\varphi = a^2/(a+b)^2$. Another important geometric parameter for calculating the critical contact angle is the ratio of the total area (Figure S3(3)) to the projected surface area, given by $r = 1 + 4ah/(a+b)^2$. Table S5. Shows $\varphi$ *and* $r$ value for the different post-spacing.

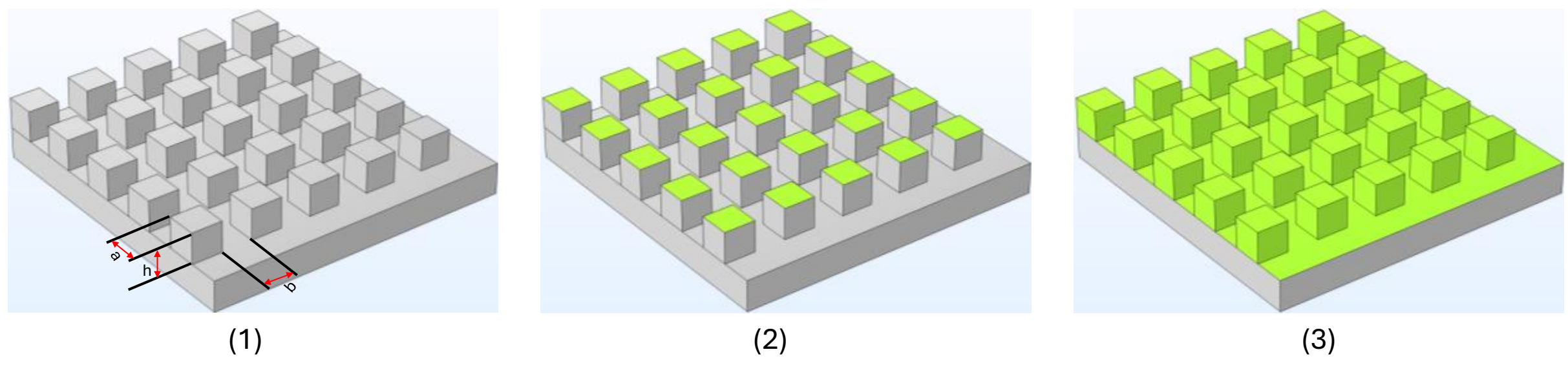

**Figure S2.** Schematic representation of square post-textured surface (1), Dimensions (2), Solid fraction (3), Total area

**Table S5.** In the case of square posts with width a, edge-to-edge spacing b, and height h, $\varphi = a^2/(a+b)^2$ and $r = 1 + 4ah/(a+b)^2$ Texture parameters *b, r,* and critical contact angles $\theta c$ defined by $\theta c = cos^{-1}((1-\varphi)/(r-\varphi))$. A schematic representation of a square post-textured surface is given in Figure S3.

| Post spacing(b)$(\mu m)$ | $r$ | $\varphi$ | $\theta c$ $(^{\circ})$ |
|---|---|---|---|
| 5 | 2.778 | 0.444 | 76.229 |
| 20 | 1.444 | 0.111 | 48.191 |

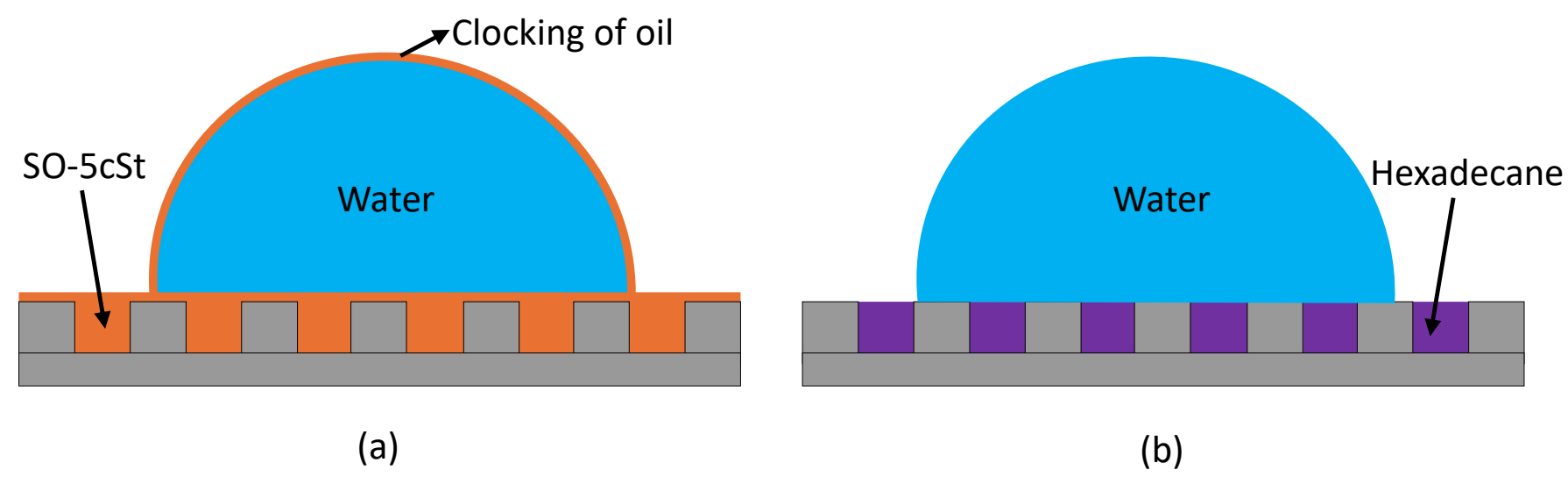

**Figure S3.** Schematic diagram of a liquid droplet placed on a textured surface impregnated with a lubricant that (a) cloaks the oil (Orange colour represents SO-5cst), (b) does not cloak the oil around the water droplet (purple colour represents hexadecane).

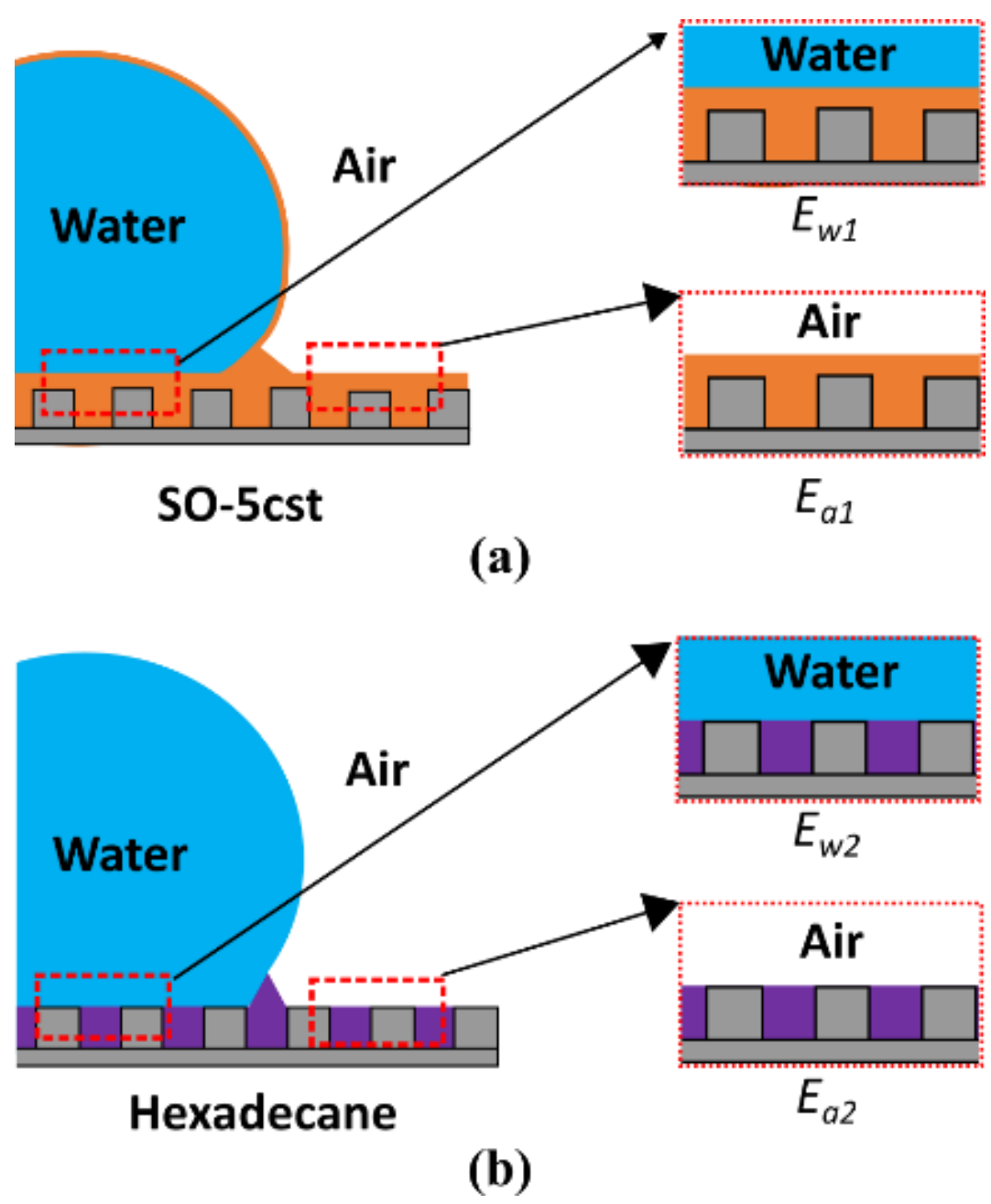

**Figure S4.** Schematics of wetting configurations drop for (a) Silicon Oil(5*cst*) and (b) Hexadecane.

**Table S6.** The total interface energy per unit area is calculated for the above configuration (Figure S5) by summing the individual interfacial energy contributions. Equivalent stability requirements for each configuration are provided in the next column.

| **Total interfacial energy per unit area accordingly to Figure S4** | **Equivalent criteria** | | |
|---|---|---|---|
| $\boldsymbol{E_{w1}} = \boldsymbol{\gamma_{wo}} + \boldsymbol{r\gamma_{os}}$ (SO-5cst) | $E_{w1} < E_{w2}$ | $S_{os(w)} \geq 0$ | $\theta_{os(w)} = 0$ |

| $\boldsymbol{E_{a1}} = \boldsymbol{\gamma_{oa}} + \boldsymbol{r\gamma_{os}}$ (SO-5$cst$) | $E_{a1} < E_{a2}$ | $S_{os(a)} \geq 0$ | $\theta_{os(a)} = 0$ |
|---|---|---|---|
| $\boldsymbol{E_{w2}} = (\boldsymbol{r} - \boldsymbol{\varphi})\boldsymbol{\gamma_{os}} + \boldsymbol{\varphi\gamma_{sw}} + (\boldsymbol{1} - \boldsymbol{\varphi})\boldsymbol{\gamma_{ow}}$ (Hexadecane) | $E_{w2} < E_{w1}$ | $-\gamma_{ow}\left(\frac{r-1}{r-\varphi}\right) < S_{os(w)} < 0$ | $\theta_{os(w)} > 0 > \theta c$ |
| $\boldsymbol{E_{a2}} = (\boldsymbol{r} - \boldsymbol{\varphi})\boldsymbol{\gamma_{os}} + \boldsymbol{\varphi\gamma_{sa}} + (\boldsymbol{1} - \boldsymbol{\varphi})\boldsymbol{\gamma_{oa}}$ (Hexadecane) | $E_{a2} < E_{a1}$ | $-\gamma_{oa}\left(\frac{r-1}{r-\varphi}\right) < S_{os(a)} < 0$ | $\theta_{os(a)} > 0 > \theta c$ |

### 3.1. Calculation of the Effective Hamaker Constant

Consider the interaction between the fluid and the PDMS substrate. Let $h$ represent the thickness of the oil layer, $d_1$ the thickness of the PDMS layer, and $d_0$ the distance between the two interfaces and the Hamaker constant is denoted by $A$. The total interaction between the substrate and fluid can be written as,

$$G^{lw}_{system} = G^{lw}_{film} + G^{lw}_{substrate} + G^{lw}_{interface}$$

$$= C_2 - \frac{A_{22}}{12\pi h^2} + C_2 - \frac{A_{22}}{12\pi d_1^2} - \frac{A_{12}}{12\pi}\left[\frac{1}{d_0^2} - \frac{1}{(d_0 + h)^2}\right]$$

$$= C_E - \frac{A_{22}}{12\pi h^2} + \frac{A_{12}}{12\pi h^2}$$

$$\therefore\ G^{lw}_{system} \sim -\frac{A_E}{12\pi h^2}$$

Where $A_E = Effective\ hamaker\ constant = A_{22} - A_{12}$

This $A_{12} = \sqrt{A_{11}A_{22}}$ form the combining relations[6]

Where, $A_{22} = 24\pi d_0^2 \gamma_2^{lw}$ form the combining relation[6]

**Given** $\gamma^{lw}_{2(SO5cst)} = Silicon\ oil\ 5cSt = 0.0197\ N/m^2$, $\gamma^{lw}_{2(hexa)} = Hexadecane = 0.027\ N/m^2$, $d_0 = 0.165\ nm$

**Table S7.** Data for the effective Hamaker constant in ($10^{-20} j$).

| | $\boldsymbol{A_{22}}$ | $\boldsymbol{A_{11}}$**(PDMS)** | $\boldsymbol{A_{12}} = \sqrt{\boldsymbol{A_{11}A_{22}}}$ | $\boldsymbol{A_E}(\times \mathbf{10^{-20}} \boldsymbol{j})$ |
|---|---|---|---|---|
| **Silicon oil** | 4.04 | 4.4 | 4.21 | -0.1801$\pm$0.150 |
| **Hexadecane** | 5.64 | 4.4 | 4.98 | 0.66$\pm$0.08 |

## 4. FESEM images of textured PDMS

The FESEM images reveal a well-defined microtextured morphology formed on the PDMS surface (Figure S5). The surface exhibits uniform, periodic microstructures with minimal defects, confirming accurate replication from the mold. At higher magnifications, fine cracks are observed on the surface. These cracks arise due to the thin (~5 nm) conductive gold coating applied to prevent charging during imaging. Overall, the microtextures are clearly captured, confirming the structural integrity of the patterned PDMS surface.

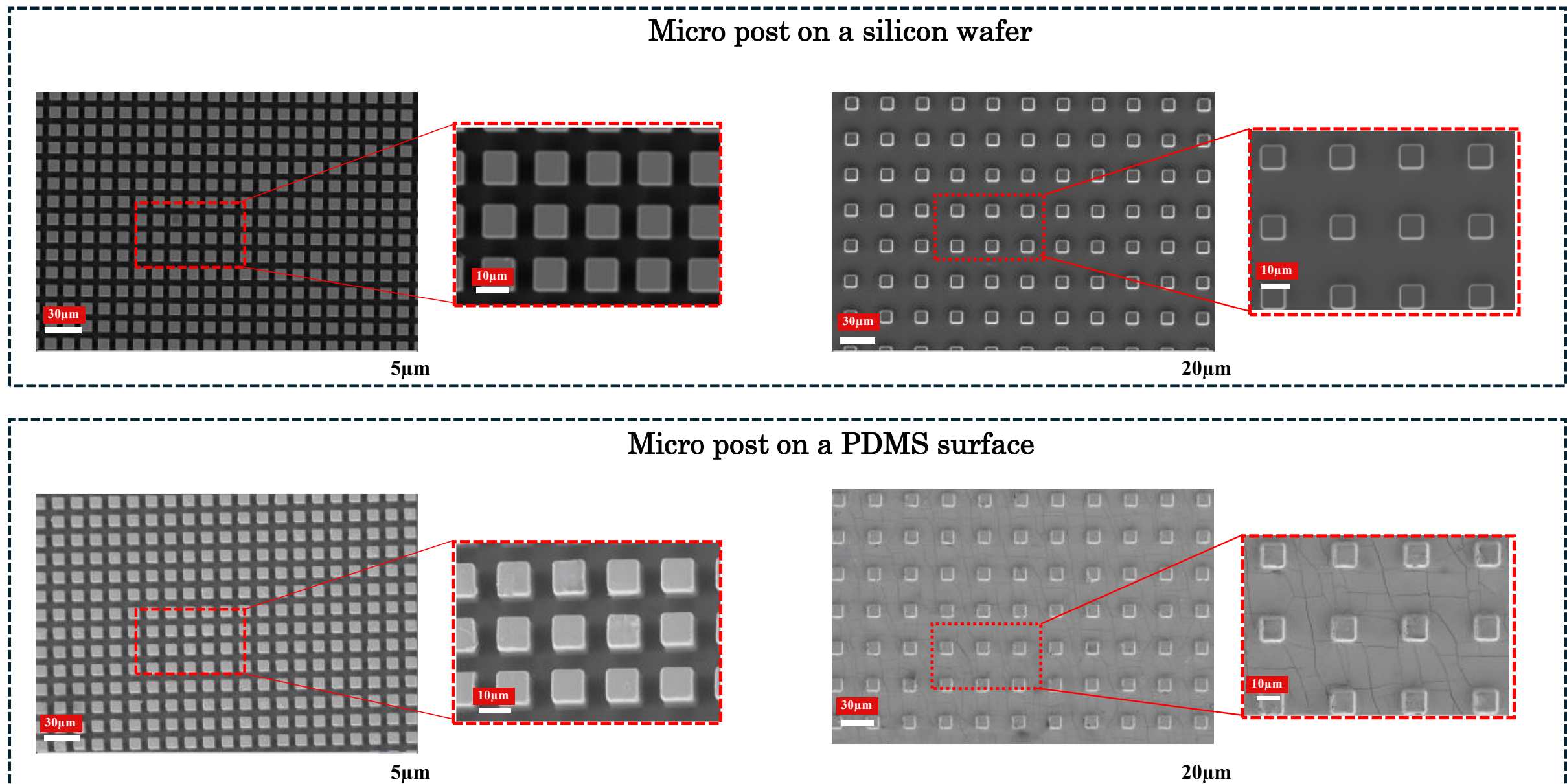

**Figure S5.** FESEM images of textured silicone wafer and textured PDMS surface at two different magnifications: for 5μm and 20μm post spacing samples, respectively.

## 5. Experimental critical We for the transition in bouncing phenomena

To determine the experimental critical (threshold) value, we conducted a series of experiments with Weber numbers of 28, 63, 127, and 247. We conducted experiments with the Weber number in small intervals (~10) to determine the nearest experimental critical value for various transitions, ranging from no rebound to partial rebound, as the Weber number increased. Table S7. Shows the values.

**Table S8.** The experimental critical values for the transition from one outcome to another.

| Sr. no | Samples | Threshold/ critical We | Outcomes |
| --- | --- | --- | --- |
| 1 | PDMS-OTS 5μm | We > ~40 | Always No Rebound |

| 2 | PDMS-OTS 20μm | - | Always No Rebound |
|---|---|---|---|
| Coated with Lubricant | | | |
| 3 | PDMS-OTS (SO-5cSt) 5μm | We ~ 220 | Partial Rebound → No Rebound |
| 4 | PDMS-OTS (Hexa) 5μm | We ~ 40 | No Rebound → Partial Rebound |
| | | We ~ 210 | Partial Rebound → No Rebound |
| 5 | PDMS-OTS (SO-5cSt) 20μm | We ~ 220 | Partial Rebound → No Rebound |
| 6 | PDMS-OTS (Hexa) 20μm | We ~ 40 | No Rebound → Partial Rebound |
| | | We ~ 210 | Partial Rebound → No Rebound |
| Absorbed with Lubricant | | | |
| 7 | PDMS-OTS (SO-5cSt) 5μm | We > 15-20 | Always Rebound |
| 8 | PDMS-OTS (Hexa) 5μm | We ~ 230 | Partial Rebound → No Rebound |
| 9 | PDMS-OTS (SO-5cSt) 20μm | We > ~ 15-20 | Always Rebound |
| 10 | PDMS-OTS (Hexa) 20μm | We ~ 30 | Partial Rebound→Full rebound |
| | | We ~ 210 | Full rebound→No rebound |

### 5.1. Calculation and explanation for wetting and anti-wetting pressure on textured PDMS-OTS surface.

The relative magnitudes of wetting and anti-wetting pressures determine the wetting states of impinging droplets[7]:

- $P_{EWH}$ is produced during the contact stage when the droplet impacts the textured surface.
- A total wetting state occurs when $P_{EWH}$ exceeds $P_D$ and $P_C$, i.e ($P_{EWH}>P_D>P_C$), allowing water to penetrate during both the contact and spreading stages.
- A partial wetting state is observed when $P_{EWH}$ is greater than $P_C$ but less than $P_D$, i.e ($P_{EWH}>P_C>P_D$), leading to water penetration only during the contact stage.
- A total nonwetting state arises when $P_C$ exceeds both $P_{EWH}$ and $P_D$, i.e ($P_C>P_{EWH}>P_D$), causing the structure to resist wetting throughout both stages.

Dynamic/kinetic pressure = $P_D = \frac{1}{2}\rho v^2$

Effective hammer pressure of water = $P_{EWH} = 0.2\rho C v$

Capillary pressure = $P_C = -2\sqrt{2}\,\gamma_{LV} \cos\frac{\theta_A}{B}$

Where, $\rho$ Density, $v$ impact velocity, $C$ velocity of sound in water=1497*m/s*, $\gamma_{LV}$ Interfacial tension of water in air =0.072 *N/m*, $\theta_A$ is advancing the contact angle of water on a smooth PDMS-OTS coated surface, and D is post-sapping.

**Table S9.** Shows the calculated values of (a) dynamic, effective hammering pressure for the particular velocity and (b) capillary pressure for the corresponding post-spacing (Measuring units are Pascal *p*).

(a)

| V(m/s)~(We) | Dynamic pressure | Hammering pressure |
|---|---|---|
| | $P_D$ | $P_{EWH}$ |
| 0.88 (28) | 387.2 | 263472 |
| 1.32 (63) | 871.2 | 395208 |
| 1.88 (127) | 1767.2 | 562872 |
| 2.61 (247) | 3406.05 | 781434 |

(b)

| Post spacing | Capillary pressure |
|---|---|
| | $P_C$ |
| 5$\mu m$ | 18658.56 |
| 20$\mu m$ | 4664.64 |

## 6. Regime map, repeatability and oil retention

**Table S10.** Values of the wettability parameter and contact angle hysteresis (CAH) for the respective samples.

| Samples | CAH | Wettabilty parameter (180/CAH) |
|---|---|---|
| 20µm Hexadecane - Coated | $30 \pm 7$ | 6 |
| 5µm Hexadecane - Coated | $28 \pm 4$ | 6.4 |
| 20µm SO-5cST - Coated | $25 \pm 6$ | 7.2 |
| 5µm SO-5cST - Coated | $24 \pm 4$ | 7.5 |
| 5µm Hexadecane - Absorbed | $12 \pm 2$ | 15 |
| 20µm Hexadecane - Absorbed | $10 \pm 2$ | 18 |
| 5µm SO-5cST - Absorbed | $1.5 \pm 1$ | 120 |
| 20µm SO-5cST - Absorbed | $1.4 \pm 1$ | 123 |

**Table S11.** Percentage (%) lubricant loss on 20µm post-spacing surfaces subjected to repeated droplet impacts at We = 127, showing the reduction in lubricant content with increasing impact until complete rebound ceases. (±)Uncertainties represent the deviation for four measurements.

| Lubricant | 1st Impact | 2nd Impact | 4th Impact | 6th Impact | 8th Impact |
|---|---|---|---|---|---|
| non-Van der Waals-SLIPS | 0.30 ±0.15 | 0.24 ±0.12 | 0.13 ±0.14 | 0.11 ±0.09 | 0.09 ±0.07 |
| Van der Waals-SLIPS | 0.10 ±0.05 | 0.10 ±0.09 | 0.05 ±0.04 | 0.03 ±0.05 | 0.02 ±0.04 |

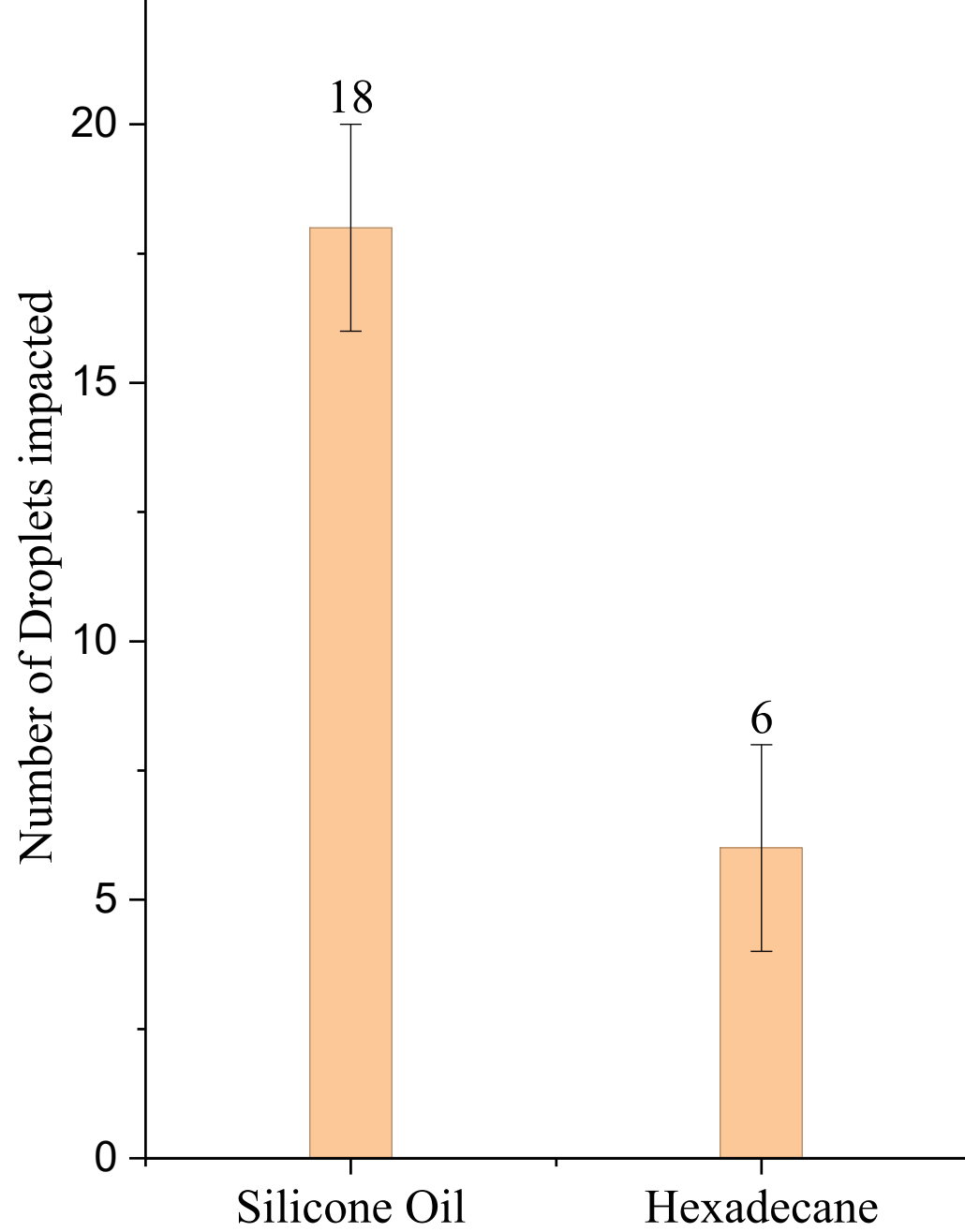

**Figure S6.** Number of droplet impacts sustained before shifting from full rebound to partial rebound on 20µm post-patterned PDMS-OTS absorbed samples using different oils.

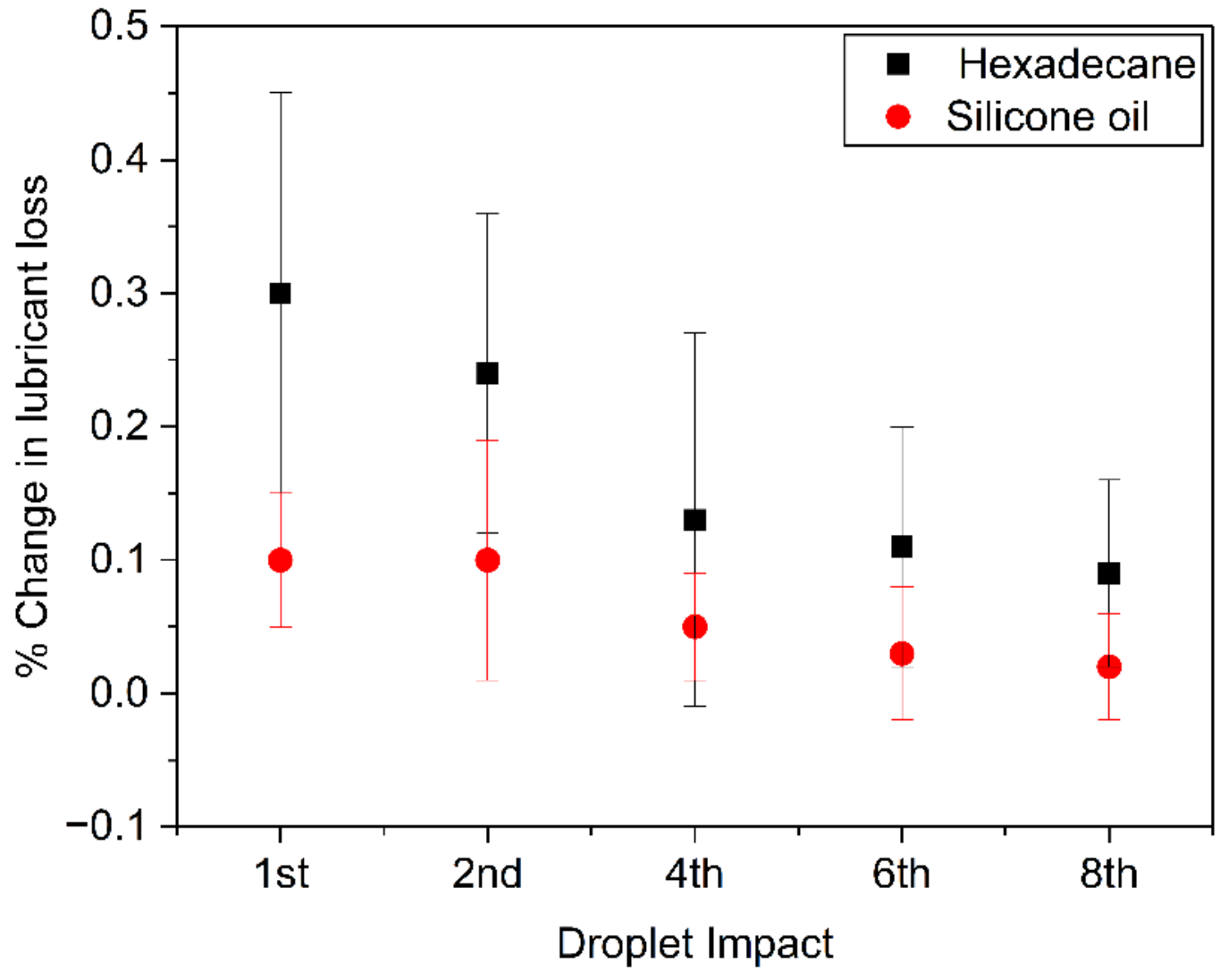

**Figure S7.** Percentage (%) lubricant loss on 20µm post-spacing surfaces subjected to repeated droplet impacts at We = 127, showing the reduction in lubricant content with increasing impact until complete rebound ceases. (±)Uncertainties represent the deviation of the four measurements

# References

(1) A. M. Worthington. On the Forms Assumed by Drops of Liquids Falling Vertically on a Horizontal Plate Author ( s ): A . M . Worthington Source : Proceedings of the Royal Society of London , Vol . 25 ( 1876 - 1877 ), Pp . 261-272 Published by : The Royal Society Stable URL : *Proc. R. Soc. London* **1876**, *25*, 261–272.

(2) Muschi, M.; Brudieu, B.; Teisseire, J.; Sauret, A. Drop Impact Dynamics on Slippery Liquid-Infused Porous Surfaces: Influence of Oil Thickness. *Soft Matter* **2018**, *14* (7), 1100–1107. https://doi.org/10.1039/c7sm02026k.

(3) Lee, E.; Chilukoti, H. K.; Müller-Plathe, F. Suppressing the Rebound of Impacting Droplets from Solvophobic Surfaces by Polymer Additives: Polymer Adsorption and Molecular Mechanisms. *Soft Matter* **2021**, *17* (29), 6952–6963. https://doi.org/10.1039/d1sm00558h.

(4) Jayaprakash, V.; Rufer, S.; Panat, S.; Varanasi, K. K. Enhancing Spray Retention Using Cloaked Droplets to Reduce Pesticide Pollution. *Soft Matter* **2025**, *21* (19). https://doi.org/10.1039/d4sm01496k.

(5) Wang, H.; Lu, H.; Zhao, W. A Review of Droplet Bouncing Behaviors on Superhydrophobic Surfaces: Theory, Methods, and Applications. *Phys. Fluids* **2023**, *35* (2), 021301 (1-23). https://doi.org/10.1063/5.0136692.

(6) Josserand, C.; Thoroddsen, S. T. Drop Impact on a Solid Surface. *Annu. Rev. Fluid Mech.* **2016**, *48* (September 2015), 365–391. https://doi.org/10.1146/annurev-fluid-122414-034401.

(7) Yarin, A. L. Drop Impact Dynamics: Splashing, Spreading, Receding, Bouncing.. *Annu. Rev. Fluid Mech.* **2006**, *38*, 159–192. https://doi.org/10.1146/annurev.fluid.38.050304.092144.

(8) Quéré, D. Leidenfrost Dynamics. *Annu. Rev. Fluid Mech.* **2013**, *45*, 197–215. https://doi.org/10.1146/annurev-fluid-011212-140709.

(9) Tran, T.; Staat, H. J. J.; Susarrey-Arce, A.; Foertsch, T. C.; van Houselt, A.; Gardeniers, H. J. G. E.; Prosperetti, A.; Lohse, D.; Sun, C. Droplet Impact on Superheated Micro-Structured Surfaces. *Soft Matter* **2013**, *9* (12), 3272–3282. https://doi.org/10.1039/C3SM27643K.

(10) Zhang, H.; Ahumada Lazo, J.; Sarwar, M. S. Bin; Liu, Y. Revealing the Dynamic and Thermal Behaviors of Supercooled Droplet Impinging on Surfaces with Varying Wettability. *Soft Matter* **2025**, *21* (45), 8655–8668. https://doi.org/10.1039/d5sm00761e.

(11) Sharma, S.; Pinto, R.; Saha, A.; Chaudhuri, S.; Basu, S. On Secondary Atomization and Blockage of Surrogate Cough Droplets in Single- And Multilayer Face Masks. *Sci. Adv.* **2021**, *7* (10), 1–12. https://doi.org/10.1126/sciadv.abf0452.

(12) Khojasteh, D.; Kazerooni, M.; Salarian, S.; Kamali, R. Droplet Impact on Superhydrophobic Surfaces: A Review of Recent Developments. *J. Ind. Eng. Chem.* **2016**, *42*, 1–14. https://doi.org/10.1016/j.jiec.2016.07.027.

(13) Yu, X.; Zhang, Y.; Hu, R.; Luo, X. Water Droplet Bouncing Dynamics. *Nano Energy* **2021**, *81* (November 2020), 105647. https://doi.org/10.1016/j.nanoen.2020.105647.

(14) Patankar, N. A. On the Modeling of Hydrophobic Contact Angles on Rough Surfaces. *Langmuir* **2003**, *19* (4), 1249–1253. https://doi.org/10.1021/la026612+.

(15) Patankar, N. A. Transition between Superhydrophobic States on Rough Surfaces. *Langmuir* **2004**, *20* (17), 7097–7102. https://doi.org/10.1021/la049329e.

(16) Nonomura, Y.; Tanaka, T.; Mayama, H. Penetration Behavior of a Water Droplet into a Cylindrical Hydrophobic Pore. *Langmuir* **2016**, *32* (25), 6328–6334. https://doi.org/10.1021/acs.langmuir.6b01509.

(17) Jung, Y. C.; Bhushan, B. Dynamic Effects of Bouncing Water Droplets on Superhydrophobic Surfaces. *Langmuir* **2008**, *24* (12), 6262–6269. https://doi.org/10.1021/la8003504.

(18) Mitra, S.; Vo, Q.; Tran, T. Bouncing-to-Wetting Transition of Water Droplets Impacting Soft Solids. *Soft Matter* **2021**, *17* (24), 5969–5977. https://doi.org/10.1039/d1sm00339a.

(19) Wang, L. Z.; Zhou, A.; Zhou, J. Z.; Chen, L.; Yu, Y. S. Droplet Impact on Pillar-Arrayed Non-Wetting Surfaces. *Soft Matter* **2021**, *17* (24), 5932–5940. https://doi.org/10.1039/d1sm00354b.

(20) Wu, J.; Zhang, L.; Lu, Y.; Yu, Y. Droplet Impinging on Sparse Micropillar-Arrayed Non-Wetting Surfaces. *Phys. Fluids* **2024**, *36* (9). https://doi.org/10.1063/5.0226032.

(21) Zhang, L.; Wu, J.; Lu, Y.; Yu, Y. Droplets Impact on Sparse Microgrooved Non-Wetting Surfaces. *Sci. Rep.* **2025**, *15* (1), 2918. https://doi.org/10.1038/s41598-025-87294-z.

(22) Villegas, M.; Zhang, Y.; Abu Jarad, N.; Soleymani, L.; Didar, T. F. Liquid-Infused Surfaces: A Review of Theory, Design, and Applications. *ACS Nano* **2019**, *13* (8), 8517–8536. https://doi.org/10.1021/acsnano.9b04129.

(23) Li, J.; Zhou, Z.; Jiao, X.; Guo, Z.; Fu, F. Bioinspired Lubricant-Infused Porous Surfaces: A Review on Principle, Fabrication, and Applications. *Surfaces and Interfaces* **2024**, *53* (July). https://doi.org/10.1016/j.surfin.2024.105037.

(24) Zhang, D.; Xia, Y.; Chen, X.; Shi, S.; Lei, L. PDMS-Infused Poly(High Internal Phase Emulsion) Templates for the Construction of Slippery Liquid-Infused Porous Surfaces with Self-Cleaning and Self-Repairing Properties. *Langmuir* **2019**, *35* (25), 8276–8284. https://doi.org/10.1021/acs.langmuir.9b01115.

(25) Li, J.; Lu, B.; Cheng, Z.; Cao, H.; An, X. Designs and Recent Progress of "Pitcher Plant Effect" Inspired Ultra-Slippery Surfaces: A Review. *Prog. Org. Coatings* **2024**, *191* (April), 108460. https://doi.org/10.1016/j.porgcoat.2024.108460.

(26) Ganar, S. S.; Das, A. Experimental Insights into Droplet Behavior on Van Der Waals and Non-Van Der Waals Liquid-Impregnated Surfaces. **2024**, No. December 2022. https://doi.org/10.1063/5.0236861.

(27) Yeganehdoust, F.; Attarzadeh, R.; Dolatabadi, A.; Karimfazli, I. A Comparison of Bioinspired Slippery and Superhydrophobic Surfaces: Micro-Droplet Impact. *Phys. Fluids* **2021**, *33* (2). https://doi.org/10.1063/5.0035556.

(28) Wong, T.; Kang, S. H.; Tang, S. K. Y.; Smythe, E. J.; Hatton, B. D.; Grinthal, A.; Aizenberg, J. Bioinspired Self-Repairing Slippery Surfaces with Pressure-Stable Omniphobicity. *Nature* **2011**, *477* (7365), 443–447. https://doi.org/10.1038/nature10447.

(29) Rapoport, L.; Solomon, B. R.; Varanasi, K. K. Mobility of Yield Stress Fluids on Lubricant-Impregnated Surfaces. *ACS Appl. Mater. Interfaces* **2019**, *11* (17), 16123–16129. https://doi.org/10.1021/acsami.8b21478.

(30) Nayse, A. K.; Mund, A.; Ganar, S.; Das, A. Design and Fabrication of Industrially Scalable Low-Cost Liquid Impregnated Omni-Phobic Surfaces with Extreme Hydratephobic Properties. *Surfaces and Interfaces* **2025**, *70* (May). https://doi.org/10.1016/j.surfin.2025.106744.

(31) Lee, C.; Kim, H.; Nam, Y. Drop Impact Dynamics on Oil-Infused Nanostructured Surfaces. *Langmuir* **2014**, *30* (28), 8400–8407. https://doi.org/10.1021/la501341x.

(32) Chen, F.; Wang, Y.; Tian, Y.; Zhang, D.; Song, J.; Crick, C. R.; Carmalt, C. J.; Parkin, I. P.; Lu, Y. Robust and Durable Liquid-Repellent Surfaces. *Chem. Soc. Rev.* **2022**, *51* (20), 8476–8583. https://doi.org/10.1039/d0cs01033b.

(33) Ganar, S. S.; J, D.; Das, A. High-Speed Imagery Analysis of Droplet Impact on van Der Waals and Non-van Der Waals Soft Oil-Infused Surfaces. *Langmuir* **2025**. https://doi.org/10.1021/acs.langmuir.5c02765.

(34) Das, A.; Farnham, T. A.; Bengaluru Subramanyam, S.; Varanasi, K. K. Designing Ultra-Low Hydrate Adhesion Surfaces by Interfacial Spreading of Water-Immiscible Barrier Films. *ACS Appl. Mater. Interfaces* **2017**, *9* (25), 21496–21502. https://doi.org/10.1021/acsami.7b00223.

(35) Ghosh, N.; Bajoria, A.; Vaidya, A. A. Surface Chemical Modification of Poly(Dimethylsiloxane)-Based Biomimetic Materials: Oil-Repellent Surfaces. *ACS Appl. Mater. Interfaces* **2009**, *1* (11), 2636–2644. https://doi.org/10.1021/am9004732.

(36) Cao, Y.; Jana, S.; Tan, X.; Bowen, L.; Zhu, Y.; Dawson, J.; Han, R.; Exton, J.; Liu, H.; McHale, G.; Jakubovics, N. S.; Chen, J. Antiwetting and Antifouling Performances of Different Lubricant-Infused Slippery Surfaces. *Langmuir* **2020**, *36* (45), 13396–13407. https://doi.org/10.1021/acs.langmuir.0c00411.

(37) Qin, D.; Xia, Y.; Whitesides, G. M. Soft Lithography for Micro- and Nanoscale Patterning. *Nat. Protoc.* **2010**, *5* (3), 491–502. https://doi.org/10.1038/nprot.2009.234.

(38) Vaillard, A. S.; Saget, M.; Braud, F.; Lippert, M.; Keirsbulck, L.; Jimenez, M.; Coffinier, Y.; Thomy, V. Highly Stable Fluorine-Free Slippery Liquid Infused Surfaces. *Surfaces and Interfaces* **2023**, *42* (PA), 103296. https://doi.org/10.1016/j.surfin.2023.103296.

(39) Dawson, J.; Coaster, S.; Han, R.; Gausden, J.; Liu, H.; McHale, G.; Chen, J. Dynamics of Droplets Impacting on Aerogel, Liquid Infused, and Liquid-Like Solid Surfaces. *ACS Appl. Mater. Interfaces* **2023**, *15* (1), 2301–2312. https://doi.org/10.1021/acsami.2c14483.

(40) Alizadeh, A.; Bahadur, V.; Shang, W.; Zhu, Y.; Buckley, D.; Dhinojwala, A.; Sohal, M. Influence of Substrate Elasticity on Droplet Impact Dynamics. *Langmuir* **2013**, *29* (14), 4520–4524. https://doi.org/10.1021/la304767t.

(41) Chen, L.; Bonaccurso, E.; Deng, P.; Zhang, H. Droplet Impact on Soft Viscoelastic Surfaces. *Phys. Rev. E* **2016**, *94* (6), 1–9. https://doi.org/10.1103/PhysRevE.94.063117.

(42) Kanungo, M.; Mettu, S.; Law, K. Y.; Daniel, S. Effect of Roughness Geometry on Wetting and Dewetting of Rough PDMS Surfaces. *Langmuir* **2014**, *30* (25), 7358–7368. https://doi.org/10.1021/la404343n.

(43) Smith, J. D.; Meuler, A. J.; Bralower, H. L.; Venkatesan, R.; Subramanian, S.; Cohen, R. E.; McKinley, G. H.; Varanasi, K. K. Hydrate-Phobic Surfaces: Fundamental Studies in Clathrate Hydrate Adhesion Reduction. *Phys. Chem. Chem. Phys.* **2012**, *14* (17), 6013–6020. https://doi.org/10.1039/c2cp40581d.

(44) Smith, J. D.; Dhiman, R.; Anand, S.; Reza-Garduno, E.; Cohen, R. E.; McKinley, G. H.; Varanasi, K. K. Droplet Mobility on Lubricant-Impregnated Surfaces. *Soft Matter* **2013**, *9* (6), 1772–1780. https://doi.org/10.1039/c2sm27032c.

(45) Israelachvili, J. N. Van Der Waals Forces between Particles and Surfaces. In *Intermolecular and Surface Forces*; Elsevier, 2011; pp 253–289. https://doi.org/10.1016/B978-0-12-391927-4.10013-1.

(46) Polytechnique, E.; Cedex, P.; Seiwert, J.; Clanet, C.; Quéré, D. Coating of a Textured Solid. *J. Fluid Mech.* **2011**, *669*, 55–63. https://doi.org/10.1017/S0022112010005951.

(47) Li, J.; Kleintschek, T.; Rieder, A.; Cheng, Y.; Baumbach, T.; Obst, U.; Schwartz, T.; Levkin, P. A. Hydrophobic Liquid-Infused Porous Polymer Surfaces for Antibacterial Applications. *ACS Appl. Mater. Interfaces* **2013**, *5* (14), 6704–6711. https://doi.org/10.1021/am401532z.

(48) Van de Velde, P.; Fabre-Parras, N.; Josserand, C.; Duprat, C.; Protière, S. Spreading and Absorption of a Drop on a Swelling Surface. *Europhys. Lett.* **2023**, *144* (3), 33001. https://doi.org/10.1209/0295-5075/ad0eed.

(49) Ganar, S. S.; Das, A. Unraveling the Interplay of Leaf Structure and Wettability: A Comparative Study on Superhydrophobic Leaves of Cassia Tora, Adiantum Capillus-Veneris, and Bauhinia Variegata. *Phys. Fluids* **2023**, *35* (11), 1–38. https://doi.org/10.1063/5.0172707.

(50) Yada, S.; Lacis, U.; Van Der Wijngaart, W.; Lundell, F.; Amberg, G.; Bagheri, S. Droplet Impact on Asymmetric Hydrophobic Microstructures. *Langmuir* **2022**, *38* (26), 7956–7964. https://doi.org/10.1021/acs.langmuir.2c00561.

(51) Guo, C.; Zhao, D.; Sun, Y.; Wang, M.; Liu, Y. Droplet Impact on Anisotropic Superhydrophobic Surfaces. *Langmuir* **2018**, *34* (11), 3533–3540. https://doi.org/10.1021/acs.langmuir.7b03752.

(52) Xu, L. Liquid Drop Splashing on Smooth, Rough, and Textured Surfaces. *Phys. Rev. E* **2007**, *75* (5), 056316. https://doi.org/10.1103/PhysRevE.75.056316.

(53) Marmanis, H.; Thoroddsen, S. T. Scaling of the Fingering Pattern of an Impacting Drop. *Phys. Fluids* **1996**, *8* (6), 1344–1346. https://doi.org/10.1063/1.868941.

(54) Gidreta, B. T.; Huang, M.; Daniel, D.; Adera, S. Drop Impact Dynamics on Hierarchically Textured Lubricant-Infused Surfaces. *Phys. Rev. Fluids* **2025**, *10* (1), 013604. https://doi.org/10.1103/PhysRevFluids.10.013604.

(55) Allen, R. F. The Role of Surface Tension in Splashing. *J. Colloid Interface Sci.* **1975**, *51* (2), 350–351. https://doi.org/10.1016/0021-9797(75)90126-5.

(56) Huang, X.; Wan, K. T.; Taslim, M. E. Axisymmetric Rim Instability of Water Droplet Impact on a Super-Hydrophobic Surface. *Phys. Fluids* **2018**, *30* (9). https://doi.org/10.1063/1.5039558.

(57) Liu, D.; Yin, H.; Wu, Z.; Luo, X. Retraction Dynamics of an Impacting Droplet on a Rotating Surface. *Phys. Rev. Fluids* **2025**, *10* (3), 033602. https://doi.org/10.1103/PhysRevFluids.10.033602.

(58) Wang, F.; Fang, T. Retraction Dynamics of Water Droplets after Impacting upon Solid Surfaces from Hydrophilic to Superhydrophobic. *Phys. Rev. Fluids* **2020**, *5* (3), 033604. https://doi.org/10.1103/PhysRevFluids.5.033604.

(59) Clanet, C.; Béguin, C.; Richard, D.; Quéré, D. Maximal Deformation of an Impacting Drop. *J.*

*Fluid Mech.* **2004**, *517*, 199–208. https://doi.org/10.1017/S0022112004000904.

(60) Bennett, T.; Poulikakos, D. Splat-Quench Solidification: Estimating the Maximum Spreading of a Droplet Impacting a Solid Surface. *J. Mater. Sci.* **1993**, *28* (4), 963–970. https://doi.org/10.1007/BF00400880.

(61) Bartolo, D.; Josserand, C.; Bonn, D. Retraction Dynamics of Aqueous Drops upon Impact on Non-Wetting Surfaces. *J. Fluid Mech.* **2005**, *545*, 329–338. https://doi.org/10.1017/S0022112005007184.

(62) Ding, S.; Hu, Z.; Dai, L.; Zhang, X.; Wu, X. Droplet Impact Dynamics on Single-Pillar Superhydrophobic Surfaces. *Phys. Fluids* **2021**, *33* (10), 102108. https://doi.org/10.1063/5.0066366.